%% file: main.tex
\documentclass{article}

\usepackage{PRIMEarxiv}

\usepackage[utf8]{inputenc} 
\usepackage[T1]{fontenc}    
\usepackage{hyperref}       
\usepackage{url}            
\usepackage{booktabs}       
\usepackage{amsfonts}       
\usepackage{nicefrac}       
\usepackage{microtype}      
\usepackage{lipsum}
\usepackage{fancyhdr}       
\usepackage{graphicx}       
\graphicspath{{media/}}     

\usepackage{xcolor}
\usepackage{enumitem}
\usepackage{multirow}
\usepackage{xcolor,colortbl}
\usepackage{booktabs}
\usepackage{array}

\usepackage{subcaption}
\graphicspath{ {./Figures/} }

\title{The Data-Expectation Gap: A Vocabulary Describing Experiential Qualities of Data Inaccuracies in Smartwatches
 }

\author{
  Dimitra Dritsa \\
  Eindhoven University of Technology, Department of Industrial Design \\
  Eindhoven\\
  \texttt{d.dritsa@tue.nl} \\
   \And
  Steven Houben \\
  Eindhoven University of Technology, Department of Industrial Design \\
  Eindhoven\\
  \texttt{s.houben@tue.nll} \\
}

\begin{document}
\maketitle

\begin{abstract}

\input{Sections/0-Abstract}

\end{abstract}


\input{Sections/1-Introduction}

\input{Sections/2-Relatedwork}

\input{Sections/3-StudyRationale}

\input{Sections/4a-OnlineReviewResults}

\input{Sections/4b-FieldStudyResults}

\input{Sections/5-Vocabulary}

\input{Sections/6-Discussion}

\input{Sections/7-Conclusion}

\section*{Acknowledgments}
We thank Thomas Wells, Olaf Adan, Hans Brombacher and Mathias Funk for their valuable feedback on the paper.

\appendix

\input{Sections/Appendix}

\bibliographystyle{unsrt}  
\bibliography{references}

\end{document}

%% file: Sections/0-Abstract.tex
Many users of wrist-worn wearable fitness trackers encounter the \textit{data-expectation gap} - mismatches between data and expectations. While we know such discrepancies exist, we are no closer to designing technologies that can address their negative effects. This is largely because encounters with mismatches are typically treated unidimensionally, while they may differ in context and implications. This treatment does not allow the design of human-data interaction (HDI) mechanisms accounting for temporal, social, emotional, and other factors potentially influencing the perception of mismatches. To address this problem, we present a vocabulary that describes the breadth and context-bound character of encounters with the data-expectation gap, drawing from findings from two studies. Our work contributes to Personal Informatics research providing knowledge on how encounters with the data-expectation gap are embedded in people's daily lives, and a vocabulary encapsulating this knowledge, which can be used when designing HDI experiences in wearable fitness trackers.

%% file: Sections/1-Introduction.tex
\section{Introduction}

Wrist-worn wearable fitness trackers, here referred to as smartwatches, are widely adopted by users interested in understanding and improving their health~\cite{Michaelis2016Wearable, karapanos2016wellbeing}. However, users often find that the reported data do not represent their experience~\cite{Liang2021SleepML,Coskun2023DataSensemaking}; for example, their device counts climbed floors while their house is flat, or they get a negative sleep score after a restful sleep. This issue is what we define in this paper as \textbf{the data-expectation gap}: people hope for data to align with their expectations, which might be an objective or subjective truth, but technology cannot always support this need. Such mismatches between data and expectations can negatively affect the engagement with self-tracking, causing tensions and mistrust~\cite{Coskun2023DataSensemaking}. Multiple studies show that encountering such discrepancies can negatively affect user experience (UX) with smartwatches~\cite{maher2017users, meyer2015designguidelines, benbunan2020user, Attig2020Abandonment, Harrison2015Barriers, Michaelis2016Wearable}, even leading to device abandonment~\cite{Coorevits2016WearableAbandonment, Lazar2015Abandonment, Harrison2015Barriers}.

Despite awareness of such discrepancies, we are no closer to designing technologies that can preempt their negative effects. This is largely because the problem of encountering discrepancies is fuzzy, starting from the lack of a clear conceptualisation. Some studies explicitly talk about data inaccuracy (e.g. ~\cite{Yang2015Accuracy}), while others talk about a mismatch between data and beliefs~\cite{rapp2016personal}, or data and feelings~\cite{Ding2021Stress}. While these concepts share an overlap, they are not fully interchangeable. For example, a mismatch between data and feelings often involves subjective qualities, such as stress, which often weakly correlate to objective data~\cite{Ding2021Stress}. Labelling these cases - where the mapping between the data and the corresponding value is not exact - as data inaccuracy may be inappropriate, while they are also conceptually different than an overestimation of steps, or other examples such as a device misidentifying an activity type because it is beyond its classification capabilities. This lack of clear conceptualisation hinders Human-Computer Interaction (HCI) researchers and practitioners from understanding the problem and systematically analysing and comparing relevant knowledge. Consequently, it becomes increasingly difficult to develop mechanisms for mitigating negative experiences arising from mismatches between data and expectations.

The ambiguity surrounding the data-expectation gap is further amplified by our limited understanding of how experiences of such discrepancies are embedded in people's daily lives~\cite{mccarthy2007technology,wright2003making, Rooksby2014LivedInformatics}. The following examples illustrate how cases involving a discrepancy between data and expectation can vary widely in context, emotional experience, outcome and implications. Fritz et al.~\cite{Fritz2014LongTerm} describe a user who earned a 30K steps badge without wearing their watch, because construction vibrations were misinterpreted as steps. This step count was clearly unrealistic, but they felt motivated to genuinely walk 5K, to get the 35K badge. While this user perceived this incident positively, discrepancies between data and expectations may also be perceived negatively in other cases, such as a user who stopped using their heart rate (HR) monitor, which they had bought for anxiety management, after hearing that the device was inaccurate~\cite{Lazar2015Abandonment}. Despite the obvious differences in context and implications in these examples, encounters of discrepancies between data and experience are still reductively framed as data accuracy issues. This lack of knowledge on the interplay between context and emergence of tensions in such encounters reduces designers' capabilities to mitigate negative experiences~\cite{Partala2011PosNegUXanalysis}, as it inhibits the customisation of interaction mechanisms for diverse contexts. These problems hinder the design of pleasurable Human-Data Interaction (HDI)~\cite{mortier2014HDI} experiences, in smartwatches and broader contexts. Overcoming these obstacles is becoming increasingly important, as smartwatches nowadays often incorporate multiple functions beyond activity monitoring~\cite{Fitbit2023Sense2}, heightening the potential for mismatches between data and expectations. We need new ways of describing the full spectrum of phenomena related to the data-expectation gap; a vocabulary that allows a shared understanding of common issues, while also drawing boundaries between different types of discrepancies and encapsulating nuances in contextual circumstances and other factors affecting the resulting experience.

Motivated by these issues, our research aims to \textit{empirically establish a vocabulary that captures the spectrum of lived experiences with the data-expectation gap, mapping their contextual characteristics and their interplay with other factors affecting the emergence of tensions in such experiences}. We envision this vocabulary as a tool for designing HDI mechanisms to mitigate tensions arising from encountering discrepancies in smartwatch data. It can also serve as an analytical tool, providing structure for analysing experiences related to the data-expectation gap. To create our vocabulary, we analysed online product reviews and conducted an in-the-wild study. The resulting vocabulary highlights contextual factors related to the data-expectation gap - temporality, activity, location, co-experience, and emotions - alongside other factors that shape users’ evaluations of mismatches, such as the history of encounters, the user's background, goals and motivations, the trust in the self and the data, and system explainability. Our study enriches Personal Informatics (PI) and HDI research by offering the following contributions: 
\begin{itemize}
    \item Empirical evidence related to factors affecting how users perceive the data-expectation gap, based on two studies.
    \item A vocabulary describing the spectrum of lived experiences related to the data-expectation gap.
\end{itemize}

%% file: Sections/2-Relatedwork.tex
\section{Related work} \label{Related work}

Users often have divergent attitudes towards data accuracy; some view it as crucial, and others are more lenient~\cite{Attig2020Abandonment}. However, the factors determining these attitudes are understudied~\cite{Attig2020Abandonment}. Here, we discuss literature relevant to mismatches between data and expectations in the context of self-tracking and PI. We reviewed works related to smartwatches and self-tracking mentioning data accuracy concerns, or deviations between data and expectations concerning estimations of behaviours, actions, states, feelings, predictions, and recommendations. We use two framings - mismatches between data and expectations \textit{as a sensemaking problem}, and \textit{as a UX problem} - as they reveal different angles of the problem, while acknowledging that there can be overlaps. Finally, we discuss the relationship between algorithmic accuracy and trust.

\subsection{Smartwatches and Self-Tracking}

Smartwatches, which nowadays have greatly evolved from simple pedometers, enable the quantification of wellbeing-related aspects, like steps and sleep quality (SQ)~\cite{Michaelis2016Wearable}. Users may purchase one for specific purposes~\cite{karapanos2016wellbeing, Harrison2015Barriers}, or exploration~\cite{karapanos2016wellbeing}. The use of smartwatches is a type of self-monitoring\cite{karapanos2016wellbeing} and has relationships with the areas of PI~\cite{Li2010StagebasedPImodel} Lived Informatics~\cite{Rooksby2014LivedInformatics} and the Quantified-Self movement~\cite{Choe2014QSPractices}. While the model of Li et al.~\cite{Li2010StagebasedPImodel} highlights technological barriers affecting smooth transitions between stages users go through when interacting with data, Rooksby et al.~\cite{Li2010StagebasedPImodel} emphasise emotional, social, and temporal aspects of self-tracking in their Lived Informatics approach. The Tracker Goal Evolution Model~\cite{Niess2018TrackerGoalmodel} further emphasises the importance of trust in the data. In this model, users have hedonic and eudaimonic needs related to qualitative goals - which are, in turn, translated to quantitative goals. Reflection and trust are important in this translation; trust in the meaningfulness of the goal, and the device's accuracy.

\subsection{Mismatches Between Data and Expectations as a Sensemaking Problem}

Within this broad landscape of self-tracking, we focus on mismatches between data and expectations - concerning data output, and not broad expectations when purchasing a device (e.g. as in the Expectation Confirmation Model ~\cite{bhattacherjee2001expectationconfirmation}). Detecting a mismatch between data and expectations during HDI can be viewed as a data sensemaking problem. Perceptions of mismatches cause tensions in the meaning-making process~\cite{Figueiredo2017fertility,Coskun2023DataSensemaking,lupton2018personal, lomborg2018temporal}, described by Lomborg et al. as a ``tension between sensory-bodily and metric knowledge''~\cite[p.4595]{lomborg2018temporal}. Users' interpretations of mismatches between data and expectations are influenced by multiple factors coming together: technological capabilities, spatial elements, senses, and affective responses~\cite{lupton2018personal}.

Users often encounter mismatches in other self-tracking contexts beyond smartwatches~\cite{Coskun2023DataSensemaking}; for example, in fertility care~\cite{Figueiredo2017fertility}, or diabetes management~\cite{Raj2019Clinicalsensemaking}. In a clinical context, Raj et al.~\cite{Raj2019Clinicalsensemaking} show that contextual and clinical factors act as ``anchors'' when users build frames. Interacting with insights defying expectations caused users to question data accuracy and increased their mistrust, leading them to question the frames and the significance of their chosen anchors. The mental models users form about device capabilities shape expectations, and when these misalign with the actual technical capabilities~\cite{shih2015use}, frustrations occur. In another study~\cite{mackinlay2013fitbitaccuracy}, previous experiences with similar technologies also fostered expectations of unreliability. Nevertheless, users may still find value in interacting with imperfect data~\cite{lupton2018personal}.

Encounters with mismatches are related to other sensemaking activities identified by Coşkun and Karahanoğlu~\cite{Coskun2023DataSensemaking}: \textit{self-calibration, data augmentation, data handling}, and \textit{realization}. In \textit{realization}, users reflect on the data following their goals and motivations, and may question the data upon encountering a mismatch, or disengage. These mismatches may be related to memory problems, unexpected findings, predictions that cannot be met, or self-perception~\cite{Coskun2023DataSensemaking}. This \textit{confrontation} may lead to tensions, data triangulation activities, or mistrust towards the data, and eventual disengagement from tracking - particularly when the tracking supports an activity with a negative undertone, such as a health issue. 

We also identified some factors which, while not explicitly discussed as such, can be connected to perceiving mismatches as a meaning-making problem. One such factor is the subjectivity of the tracked parameter, often combined with a lack of knowledge of potential differences between physiological and psychological measurements corresponding to the same phenomenon~\cite{Ding2021Stress}. Data types related to emotions, such as electrodermal activity (EDA), are sometimes even interpreted in different ways (e.g., stress or excitement) by different companies~\cite{Hollis2018DataEmotions}. The unit of analysis also matters. Users may form mental models of cause-and-effect relationships based on a large stream of events, while data insights may only focus on a specific subset. For example, a user expected to see in their data a relaxed sleep after having a relaxed day, and when this expectation was disconfirmed, they felt that the data did not reflect their experience~\cite{smith2017health}. Users' activities to assess data accuracy could also be seen as a sensemaking task. Yang et al.~\cite{Yang2015Accuracy} describe two approaches; ``ad-hoc assessment'' - intuitively identifying patterns of alignment between data and experience - and ``folk-testing'', which involves systematic comparisons with ground-truth data or similar devices. However, these approaches often lead to problematic accuracy assessments~\cite{Yang2015Accuracy}. Users also often do not understand algorithm behaviour or the meaning of measurements, which hinders making sense of mismatches. The user's information-seeking needs, determined by their expertise and context, also matter. Amateur athletes may prioritise accuracy, while elite athletes value more practical, physical and social comfort~\cite{Rapp2020Eliteathletes}.

\subsection{Mismatches Between Data and Expectations as a UX Problem}

Another way to examine encounters with mismatches involves focusing on how users \textit{experience} them. UX highlights subjective qualities of experience, like emotions~\cite{Lallemand2015UXSurveyConsensus, Agarwal2009UXEmotion, Partala2011PosNegUXanalysis}. Recent research highlighted the importance of understanding experiential aspects of self-tracking in sports~\cite{postma2024SportsDataExperience}. Although no study solely focuses on experiential aspects of encounters with data inaccuracies, several studies have examined factors leading to positive or negative experiences with smartwatches~\cite{Michaelis2016Wearable, ryan2019anxious, karapanos2016wellbeing}. Evidence shows that smartwatches are associated with more positive than negative experiences~\cite{ryan2019anxious, karapanos2016wellbeing, Choudhury2021WearableDistress}, and fulfil needs related to autonomy, competence, and physical thriving~\cite{karapanos2016wellbeing}. Still, data inaccuracies often frustrate users~\cite{maher2017users, meyer2015designguidelines, benbunan2020user, Attig2020Abandonment, Harrison2015Barriers, Michaelis2016Wearable} - which indicates the presence of strong emotions associated with this experience - leading to potential device abandonment~\cite{Coorevits2016WearableAbandonment, Lazar2015Abandonment, Harrison2015Barriers}. Data accuracy is also associated with positive experiences when present~\cite{Michaelis2016Wearable}. 

Social interaction, an important factor in UX~\cite{Roto2011UXWhitePaper, Lallemand2015UXSurveyConsensus}, may also influence experiences with mismatches. Ding et a. discuss it as a mechanism that could enhance data interpretation by sharing information and experiences~\cite{Ding2021Stress}, and Jiang et al.~\cite{Jiang2023Intimasea} further explore this, showing that collaborative sensemaking can mediate initial perceptions of mismatches.

UX is also highly context-dependent\cite{Law2009ScopingUXdefinitions, Partala2011PosNegUXanalysis, karapanos2016wellbeing} and affected by temporal dynamics\cite{Lallemand2015UXSurveyConsensus, Karapanos2009UXTime, mccarthy2007technology}. A study of an early Fitbit model~\cite{mackinlay2013fitbitaccuracy} found that users noticed inaccuracies in specific contexts, like erroneous distance measurements during treadmill running, or overestimated steps during Zumba. Yang et al.~\cite{Yang2015Accuracy} showed that differences in users' activities affected their expectations and perceptions of accuracy. The broader context also mattered; while users tolerated consistently appearing inaccuracies, accuracy mattered when concerning health issues and when it made an exercise protocol ineffective or harmful. Pantzar et al.~\cite{Pantzar2017SituatedObjectivity} introduce the concept of `situated objectivity' in self-tracking to highlight the context-dependent nature of assessing measurement accuracy and value. They find that encounters with unexpected or contradicting data can prompt users to self-reflect about the meaning of data and what constitutes a healthy life for them.

\subsection{The Relationship Between Algorithmic Accuracy and Trust}

Multiple studies show that perceiving mismatches between data and expectations can lead to mistrust~\cite{Ding2021Stress, Coskun2023DataSensemaking, rapp2016personal}. Bodily awareness can considerably influence trust. In the context of exercise, Lomborg et al.~\cite{lomborg2018temporal} show that users with heightened bodily awareness trust their feelings over device data, using the device mainly to confirm their assumptions. Users may also sometimes trust algorithmic outputs more than their perception. This has been shown in other contexts, e.g. on the perception of personal trait profiles generated from social media data~\cite{Warshaw2015ReactiontoPersonalityData}. Moodlight - an installation~\cite{Snyder2015Moodlight} where the light colour changes based on the user's arousal levels - is an example of a system participants overtrusted to understand their feelings. Hollis et al.~\cite{Hollis2018DataEmotions} showed users two different framings of the same EDA data: one negative, indicating stress, and one positive, indicating engagement. The negative framing increased stress, while the positive framing reduced it. They attributed this tendency to trust incorrect system output to self-awareness issues and limited understanding of the system. Users may sometimes even trust random algorithmic output, as demonstrated in an experiment where users found plausible the random evaluations of a system assessing their mood~\cite{springer2017dice}. Another study~\cite{Costa2016Falsefeedback} found that false feedback suggesting a lower HR during a stressful task helped reduce users’ anxiety. The relationship between accuracy and trust is, therefore, complex.

\subsection{Knowledge Gaps} \label{Knowledge gaps}

\begin{table}[h]
    \centering
    \begin{tabular}{|>{\raggedright\arraybackslash}p{0.45\linewidth}|>{\raggedright\arraybackslash}p{0.45\linewidth}|} \hline 
         \rowcolor{lightgray} \textbf{Factors influencing the emergence and perception of mismatches}& \textbf{Effects}\\\hline 
         Expectations of device capabilities&Cause users to be cautious~\cite{mackinlay2013fitbitaccuracy}, or cause frustrations when high expectations are not matched~\cite{Consolvo2008Ubifit,shih2015use}\\ \hline 
         Subjectivity of tracked parameter and data literacy&Lead to wrong expectations for alignment between data and experience ~\cite{Ding2021Stress} \\ \hline 
         Unit of analysis&Leads to wrong expectations for alignment between data and experience ~\cite{smith2017health} \\ \hline 
         Testing protocols&Lead to problematic assessments of accuracy~\cite{Yang2015Accuracy} \\ \hline 
         Expertise (exercise context)&Affects information seeking needs~\cite{Rapp2020Eliteathletes} \\ \hline 
         Social interaction&Enhances data interpretation by information sharing~\cite{Ding2021Stress,Jiang2023Intimasea} \\ \hline 
         Context&Can cause differences in the degree and perception of inaccuracies that users encounter~\cite{mackinlay2013fitbitaccuracy,Yang2015Accuracy, Pantzar2017SituatedObjectivity} \\ \hline 
         Trust&Trust in the technology can be negatively affected by encounters with mismatches~\cite{rapp2016personal}. Trust in the body may determine how the users are affected~\cite{lomborg2018temporal} \\ \hline
    \end{tabular}
    \caption{Factors influencing the emergence and perception of mismatches, and their effects}
    \label{tab:Review_table}
\end{table}

Our analysis showed that the discrepancies users perceive in data include differences between data and expectations, sensations, beliefs, or feelings. We encountered various terms used to describe these discrepancies: ``mismatch between data and self-trackers’ expectations''~\cite{Coskun2023DataSensemaking}, ``contradictory or ambiguous results''~\cite{Figueiredo2018fertilityemotions}, ``discrepancies between what the data tell them about their bodies and selves and their own sensory perceptions''~\cite{lupton2018personal}, ``counterintuitive insights''~\cite{Raj2019Clinicalsensemaking}, ``perceived accuracy'' \cite{Jiang2023Intimasea, rapp2016personal}, ``discrepancy between some data automatically collected by the tools and what they believed about their behaviors''~\cite{rapp2016personal},`` `evidence' that contradicts their bodily sensations''~\cite{lomborg2018temporal}. The lack of a shared vocabulary hinders understanding commonalities and differences in these studies' references to discrepancies. The term ``data inaccuracy'' would also not suit all these cases, as it is associated with the notions of trueness and precision~\cite{Yang2015Accuracy, ISO2024} which may not apply to all these cases. Users may characterise devices as inaccurate when data do not meet their expectations; however, their perceptions are often based on faulty testing protocols and limited knowledge about device capabilities~\cite{Yang2015Accuracy}.

Beyond the lack of a common vocabulary, most studies broadly describe issues related to mismatches without examining how different types of discrepancies are perceived. Table~\ref{tab:Review_table} summarises our review's findings on factors influencing the emergence and perception of mismatches. Among these findings, only some refer to smartwatches, as others refer to general self-tracking practices~\cite{smith2017health,Pantzar2017SituatedObjectivity} or various types of PI tools~\cite{Consolvo2008Ubifit,Rapp2020Eliteathletes, rapp2016personal, lomborg2018temporal}. Information relevant to mismatches is fragmented in different studies. There is a lack of studies examining in detail how users perceive mismatches, except Yang et al.~\cite{Yang2015Accuracy} who, though, examine data accuracy from the perspective of user evaluation strategies. Knowledge of experiential aspects of mismatches is notably lacking. Despite indications that the context matters, there is a lack of a detailed analysis of different contextual characteristics and their impact on experiences of mismatches. The relationship between the user's emotional state, mismatches, and their perception is also unclear. An early study of an activity sensing system~\cite{Consolvo2008Ubifit} illustrated how seemingly similar error types can engender different experiences due to varying expectations. Seeing how responses to errors might differ, Consolvo et al.~\cite{Consolvo2008Ubifit} note that \textit{`` traditional error metrics are not the most helpful way to describe the effectiveness of such activity inference systems, but rather a new terminology is needed that considers the subtleties of the user’s perspective and how they react to the different types of errors.''} Such a vocabulary remains still lacking.

%% file: Sections/3-StudyRationale.tex
\section{Study rationale}

\subsection{Problem Statement}
Following the knowledge gaps we identified in Section~\ref{Related work}, our goal is to create a vocabulary that consolidates information relevant to the emergence and perception of mismatches between data and expectations, and brings attention to the breadth and the context-bound character of encounters with such mismatches. To create this vocabulary, we study the types of mismatches users encounter, the context surrounding them, and other factors affecting how users perceive them.

\subsection{Defining the Data-expectation Gap}

To avoid misinterpretations regarding the ``correctness'' of the reference value users compare with data, we introduce the term ``data-expectation gap'' as an umbrella term covering the descriptors we presented in Section (~\ref{Knowledge gaps}). It describes \textit{a mismatch between the detected and the expected value, related to user behaviours, actions, sensations, beliefs, or feelings, including classification and numeric value detection or estimation, or irrelevant recommendations}. This term emphasises that we focus on the perception of a deviation from expectations, rather than the truthfulness of the user's evaluation. These expectations specifically concern expectations regarding the data reported by the device. The data-expectation gap, as we conceptualise it, can cover both objective and subjective ``truths.'' For example, it can include the perception of a difference between actual and detected steps, or between self-reported and detected stress. We refer to a single instance of the data-expectation gap as a \textit{mismatch}.

\subsection{Approach}

To understand the data-expectation gap, we consider all kinds of mismatches in our analysis excluding those caused by a continuous device malfunction that is not resolved relatively quickly.  Our approach is influenced by the Technology as Experience (TaE) framework~\cite{mccarthy2007technology, wright2003making}, which emphasises the importance of understanding the richness and complexity of experiences in the wild~\cite{rogers2011wild}. We chose this framework as our analysis of previous works showed that there is much to gain by viewing this problem using a sensemaking as well as a UX framing, and the TaE framework conceptualises experience by unifying these perspectives. It highlights the compositional, sensual, emotional, and spatio-temporal threads characterising experiences, and discusses how the perception and meaning of experiences may change when making sense of an experience - when anticipating it, interpreting it, or reflecting on it. We draw on these themes in our analysis, while also allowing our own interpretations to surface.

\begin{figure*}[h]
    \centering
    \includegraphics[width=0.7\textwidth]{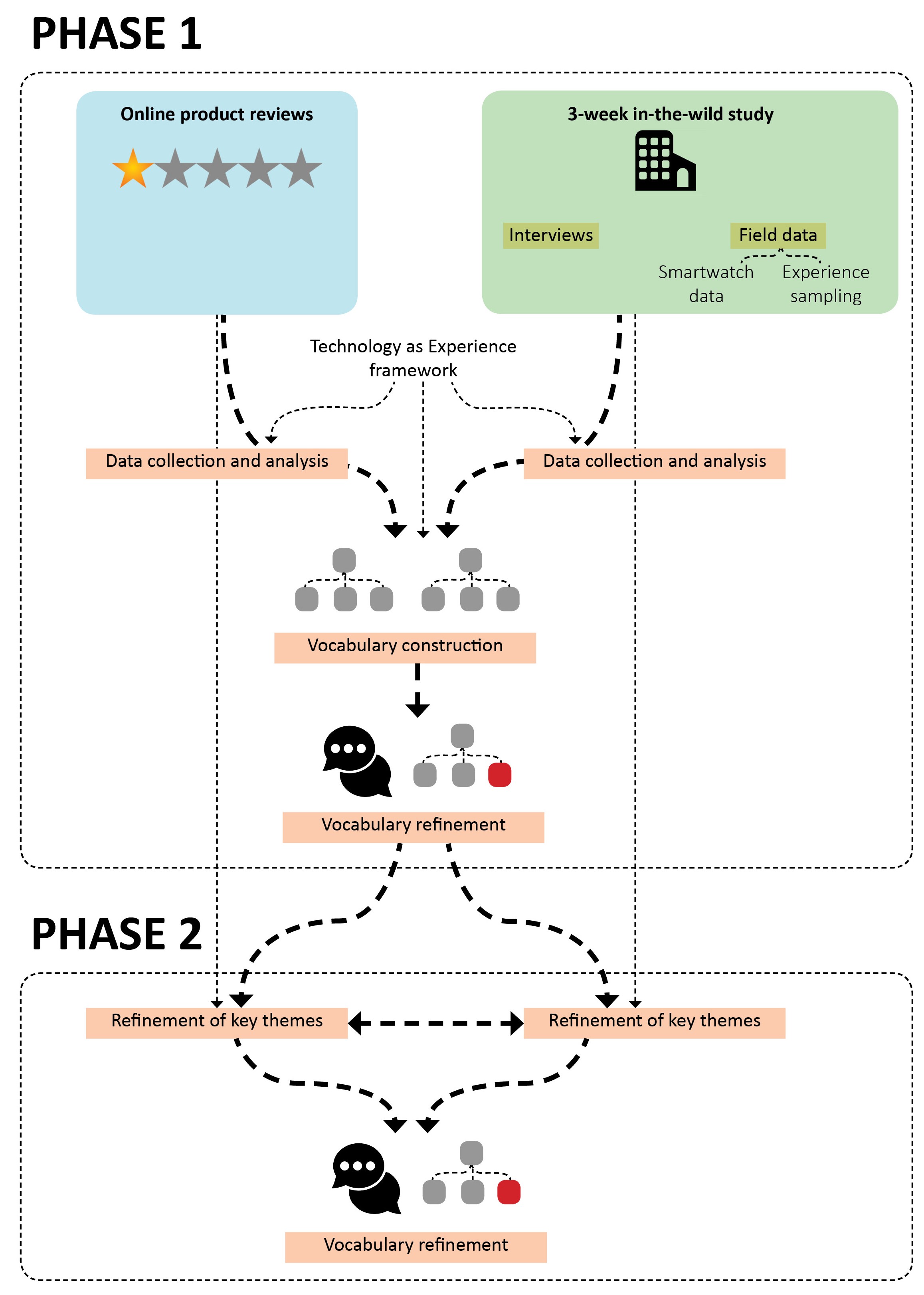}
    \caption{The process we followed to construct the vocabulary.}
    \label{fig:methodoverview}
\end{figure*}

To establish our vocabulary, we analysed three data types collected in two studies: online product reviews, interviews, and field data. In the works reviewed in Section \ref{Related work}, qualitative data proved essential for understanding felt experiences~\cite{meyer2015designguidelines,Yang2015Accuracy}; however, online product reviews are also useful for revealing broad patterns in UX~\cite{Yang2015Accuracy, benbunan2020user}. Detailed logs of errors experienced in the wild can also complement users' recollections of related experiences~\cite{Consolvo2008Ubifit}, corroborating them or providing new angles. Our study, which was approved by the university ethics board (\textit{number anonymised for review}), follows a mixed-methods design that combines the strengths of these approaches. The process (Figure~\ref{fig:methodoverview}) was as follows:
\begin{itemize}
    \item \textbf{Phase 1}
    \begin{itemize}
        \item Collection and analysis of online review data
        \item Collection and analysis of field and interview data from the in-the-wild study
        \item Creation of the first iteration of the vocabulary by grouping the key themes found in one or both studies in overarching categories
        \item Refinement of the first iteration after discussion 
    \end{itemize}
    \item \textbf{Phase 2}
    \begin{itemize}
        \item Iterative refinement of key themes of each study, based on input from the vocabulary and themes from the TaE framework
        \item Refinement of the vocabulary terms and structure 
    \end{itemize}
\end{itemize}

The first author analysed the data and created the first iteration of the vocabulary, while both authors were involved in subsequent refinements of the vocabulary. We first present the two studies and then we present an overview of the vocabulary. We also consider the \textit{data-expectation gap} a part of our vocabulary, using the terms \textit{detect} and \textit{perceive} the data-expectation gap or a single mismatch to distinguish the step of noticing a discrepancy (\textit{detect}) from that of assigning a specific meaning to it (\textit{perceive}). With the term \textit{experiencing} the data-expectation gap or a single mismatch, we refer to the broader lived experience surrounding it, including aspects such as contextual and emotional circumstances. We refer to \textit{qualities} of the data-expectation gap to describe all factors characterising felt experiences.

\subsection{Positionality}

Both authors are white European and able-bodied, with one identifying as female and one as male. This study was motivated by shared experiences in analysing data from wearables and comparing them to subjective experiences. The lead author is also a long-term user of smartwatches, still using one despite several experiences with the data-expectation gap. These experiences and her knowledge of the device's capabilities biased the generation of the scenarios of mismatches discussed in the interviews, as personal anecdotes were shared during the discussions. While the second author also had experiences with smartwatches, he had a more unbiased perspective as he never owned a smartwatch. Our position is, therefore, formed by a blend of insider and outsider perspectives~\cite{holmes2020positionality}.

%% file: Sections/4a-OnlineReviewResults.tex
\section{Study 1 - Online Product Reviews}

\subsection{Methods}

\subsubsection{Data Collection}

\begin{figure*}
\centering
\begin{minipage}{1\textwidth}
  \centering
  \captionsetup{width=1\linewidth}
  \includegraphics[width=0.88\textwidth]{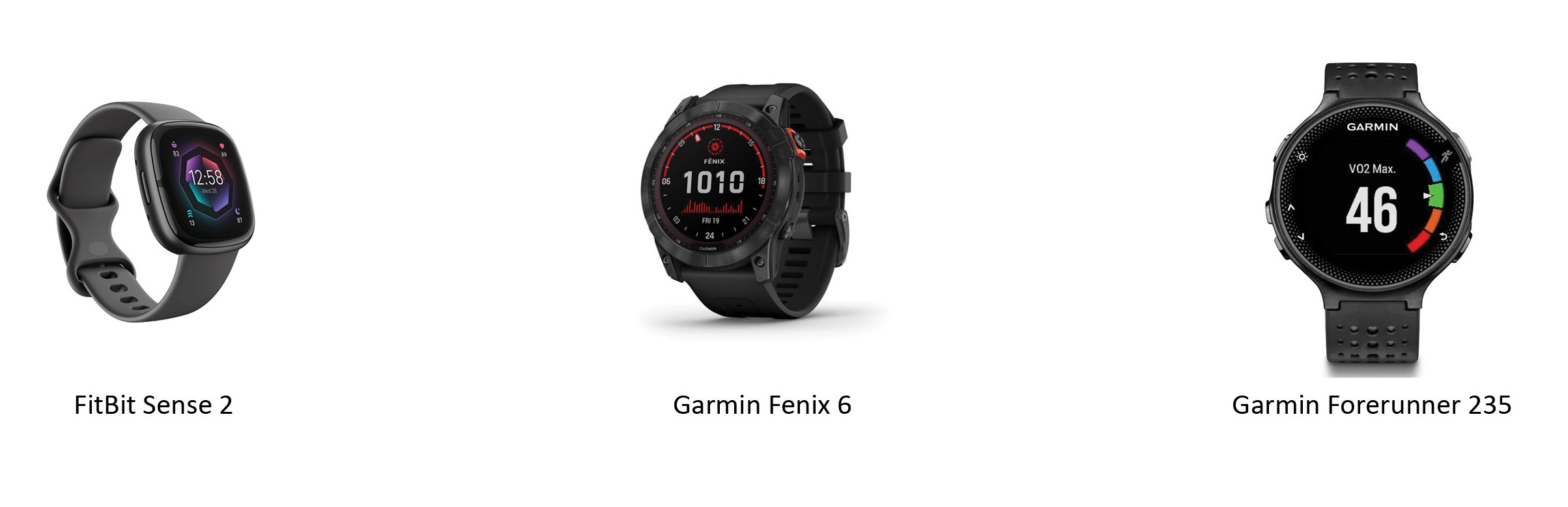}
  \captionof{figure}{The smartwatches that we considered in the detailed analysis of online reviews. Fitbit Sense, which is not shown in the image, is very similar in appearance to Fitbit Sense 2. The figures were retrieved by Fitbit~\cite{Fitbit2023} and Garmin~\cite{Garmin2023}.}
  \label{fig:DevicesOR}
\end{minipage}%
\vspace{-1.5em}
\end{figure*}

The first study was our first effort towards understanding experiences with the data-expectation gap. We were interested in mapping the types of mismatches users encounter and factors affecting how users perceive them. We sourced product reviews from Amazon, choosing 200 reviews for analysis. Details related to selecting this sample are provided in \ref{Appendix: Review selection}. The reviews covered four models: Fitbit Sense and Sense 2, Garmin Forerunner 235, and Garmin Fenix 6 (Figure \ref{fig:DevicesOR}). The sample included reviews with 1-5 stars, as sometimes participants gave high ratings despite mentioning accuracy issues. Based on our initial screening (see \ref{Appendix: Review selection}) the final sample of 200 reviews was sufficient for understanding the breadth of accuracy issues users experience. 

\subsubsection{Data Analysis}
One researcher coded the data using software for qualitative data analysis (NVivo), and then refined the codes after iterative analysis and discussions with the second author until consensus. Our analysis included a mix of inductive and deductive coding. Details are provided in \ref{Appendix: review coding protocol}.

\subsection{Results}

In the 200 analysed reviews, accuracy was the primary or only concern for 45\% of the participants, and for 41\% it co-existed with other problems, like connectivity issues, premium subscription costs, delayed reporting, charging technology, and other parameters. These reviews, where participants evaluated negatively their experiences with the data-expectation gap, typically had ratings between 1-3 stars. The rest accepted the perceived mismatches, giving ratings between 3-5 stars. 

Device use patterns varied considerably. Some participants only used specific features at specific moments, such as HR during workouts, driven by individual goals and interests. Many participants admitted unfamiliarity with certain features, having not yet used them. 

Strategies participants employed to address the data-expectation gap also varied. Some participants (10\%) contacted support, especially for pricier models, while others adopted workarounds (12\%), such as adjusting or cleaning the band, pairing the device with other devices such as HR monitoring straps, and manually editing the data.

\newcommand{\TDataLogic}{data-logic mismatch}
\newcommand{\TDataMeasure}{data-measurement mismatch}
\newcommand{\TDataFeeling}{data-feeling mismatch}
\newcommand{\TRepeatedMeasures}{inconsistency in repeated experiences}

\textbf{Source of evidence} - We identified three themes describing the sources of evidence participants compared with the data when detecting the data-expectation gap: \textit{\TDataLogic} (55\%), \textit{\TDataMeasure} (29\%), and \textit{\TDataFeeling} (10\%). Some experiences spanned multiple themes.

\begin{itemize}
    \item \textbf{\TDataLogic} - Participants often found the data implausible due to defying their assumptions of how the data should look based on their knowledge, beliefs and logic. Participants often characterised data related to such mismatches using descriptors such as ``wildly off'', ``rubbish'', ``insane'', or ``utopian'', highlighting how this category relates to a high certainty that the data is false, and therefore, cannot be trusted. Sometimes, the mismatch was such that the data clearly defied reality; for example, the device counted floors while participants were running on flat terrain, or even counted steps while the participant was using a wheelchair. There were also cases where participants were convinced that the data were implausible, but the truth of their assessment was not always clear. For example, disconfirmed expectations that HR would increase much during exercise led to perceptions of errors. In these cases, the mismatch might be due to sensor issues, but also due to erroneous mental models of physiological functions.
    \item \textbf{\TDataMeasure} - Here, participants perceived mismatches through cross-referencing data with other sources to gauge accuracy. Participants often became disappointed after finding mismatches between the HR data output of their smartwatch and other sources - their treadmill, their typically more accurate chest strap, or their manually measured pulse. Participants also tended to use for comparison other working watches that they owned or borrowed. Mistrust arose particularly when the other source was more reliable - for example, when participants found the distance completed in an outdoor route contradicting with map information. 
    \item \textbf{\TDataFeeling} - These mismatches are related to the sensual and emotional thread of experience~\cite{mccarthy2007technology}. Here participants found data contradicting their feelings or perceived state, like their stress. They reported, for example, that the watch always gave them very positive stress ratings, even though they had anxiety. Strong bodily experiences were sometimes used as evidence supporting their feeling, such as feeling the heart pumping intensely, or breathing fast. This theme sometimes overlapped with the \textit{\TDataLogic}, with the bodily experiences acting as evidence of the data implausibility - for example, participants could not believe the low HR that their watch showed, as they would feel their body sweating and their heart beating fast during training, or ``huffing and puffing'' while climbing steep terrain. 
\end{itemize}

\textbf{Type of mismatch} - Mismatches were related to parameters such as HR, steps, GPS, stress, activity classification, SpO2, speed, sleep, floor count, altitude, distance, detection of atrial fibrillation, calories, readiness or energy levels, and active minutes. Some participants also noted wrong energy-level recommendations stemming from other mismatches. Participants described the following types of mismatches:
\begin{itemize}
    \item \textbf{Classification issues:}
    \begin{itemize}
        \item \textit{False detection} (7\%) - This category is related to the identification of a \textit{false positive}. Reported examples included wrongly detecting activity while participants were inactive, and falsely identifying sleep while participants were awake. Most instances in this category related to the \textit{\TDataLogic}. 
        \item \textit{Detection failure} (7\%) - This category is related to the identification of a \textit{false negative}. It includes instances of the \textit{\TDataLogic} ~and the \textit{\TDataFeeling} - failures to detect episodes of arrhythmia in the electrocardiogram (ECG) data, episodes of stress, asleep state, and the correct activity; for example, a participant reported that the watch failed to detect walking intervals during running, resulting in wrong statistics.
        \item \textit{Partial detection} (8\%) - In sleep and activity data, the device sometimes detected an overall correct state but made mistakes in details or derivative parameters. Sleep time was sometimes measured intermittently, especially for participants whose sleep was interrupted by prolonged awakenings. Similarly, the device sometimes detected correctly that an activity started, but stopped it prematurely after detecting a pause (e.g. at traffic lights).
    \end{itemize}
    \item \textbf{Value estimation issues:}
    \begin{itemize}
        \item \textit{High fluctuation} (5\%) - A few participants noted that the values fluctuate too much - typically concerning running speed, with GPS inconsistencies presented as a possible explanation. 
      
        \item \textit{Overestimation} (33\%) - Many participants reported that the device overestimated their HR values under various conditions, including exercise, rest and sleep. Steps were also frequently exaggerated, as well as floors. Less frequent overestimations concerned calories, stress index, readiness, swimming distance, speed, and VO2max. In some cases, the conditions were such that the data should be zero, leading to particular surprise or frustration when participants saw any positive values. This happened when participants fell asleep before midnight, and woke up seeing a value larger than zero in the steps or floors, despite not having moved; or, when HR values were measured when they were not wearing the device. The presence of positive values in these cases acts as clear evidence for the \textit{\TDataLogic}.
        \item \textit{Underestimation} (28\%) - Here, most participants reported an underestimation of their HR values,  distance, and steps.
    \end{itemize}
    \item \textbf{Others:}
    \begin{itemize}
        \item \textit{Delayed response} (4\%) - Here, participants mentioned a lag in seeing the ``correct'' data. This typically happened with HR data at the beginning of exercise. Participants mentioned this issue as often frustrating but eventually perceived it as resolved. 
        \item \textit{Incorrect location} (7\%) - This category concerned GPS errors, which participants mentioned as standalone problems (e.g. during hiking), or as the reason for other mismatches, such as issues in the resulting speed or distance.
    \end{itemize}
\end{itemize}

From here onwards, we dive into how these \textit{types of mismatches} gain different meanings when combined with contextual qualities. We describe how perceptions of the data-expectation gap are influenced by the following contextual elements: \textit{activity}, \textit{location}, \textit{temporality}, \textit{personal context}, and \textit{social dimension}.

\textbf{Influence of activity and location in experiences with the data-expectation gap} - Participants engaged in various activities in their encounters with the data-expectation gap. Most experiences (65\%) were associated with exercise, such as walking, running, swimming, or hiking, while 15\% were associated with everyday activities, like working, watching TV, driving, shopping, or cooking, and 9\% were associated with rest. Everyday activities were almost exclusively associated with the \textit{\TDataLogic}~(14\%), while the context of exercising was primarily associated with the \textit{\TDataLogic}~(28\%) and the \textit{\TDataMeasure}~(23\%), and secondarily with the \textit{\TDataFeeling}~(7\%). The experiences also concerned various types of environments: 19\% referred to indoor environments, among which half concerned the participant's home, and 26\% concerned urban or natural outdoor environments. Indoor environments were mostly associated with the \textit{\TDataLogic}~(27\%), while outdoor environments were equally associated with the \textit{\TDataLogic}~and the \textit{\TDataFeeling}. Purposeful tests for comparing different measurements were more frequent during outdoor exercise, and rare in everyday activities and indoor environments.

While not all reviews mentioned details regarding the activities and locations surrounding mismatches, there were clear links between these two contextual factors and participants' perceptions of the data-expectation gap. Within each category, the factors affecting how participants perceived the data-expectation gap were the following:
\begin{itemize}
    \item \textbf{Activity}: \textit{activity type}, \textit{activity intensity},
    \item \textbf{Location}: \textit{environmental affordances and equipment}, including presence of nature, coverage, building layout and terrain.
\end{itemize}
These factors played the following roles: \textit{causing or amplifying disbelief}, \textit{causing concern, disappointment or frustration as a result of the data-expectation gap}, and \textit{causing the data-expectation gap}.

\textit{Causing or amplifying disbelief} - Participants' data expectations were closely tied to the nature of their activities. Mismatches between data and expectations \textit{in specific locations or activity contexts} often amplified perceptions of implausibility. For instance, participants frequently reported step overestimation during non-walking activities such as driving a tractor, cycling, stirring the oatmeal, using a riding lawn mower, or even sitting on the couch. The non-walking nature of the \textit{activity type} was the factor generating the belief that the data were unreliable. The environmental affordances, such as the building layout, often contributed to this perception; for example, participants reported unrealistic floor counts in flat apartments, or overestimated steps during prolonged desk work in an apartment that was too small to make this result believable. Such experiences were typically related to the \textit{\TDataLogic} and eroded trust in the device.

\textit{Causing concern, disappointment or frustration} - The combination of \textit{activity}, \textit{location}, and data parameter often triggered secondary emotional reactions in response to the data-expectation gap. When participants perceived their \textit{activity intensity} as high but the data disagreed, feelings of disappointment or frustration emerged, particularly when HR was underestimated during intense workouts. Environmental factors, such as terrain type, also influenced these emotions. Frustration or disappointment, for example, was high when walking or exercising on terrains such as stairs, a hill, or a sloped treadmill, failed to increase HR as expected. The specific emotion depended on the combination of conditions. Changes in \textit{activity type} or \textit{location} often resulted in different experiences and implications for the same data parameter. GPS-related mismatches exemplified this: during runs or walks on familiar routes, mismatches in expected and reported distance or speed due to GPS errors led to frustration or disappointment. In contrast, hikers relying on GPS in unfamiliar terrain experienced feelings of unsafeness or concern when GPS errors occurred.

\textit{Causing the data-expectation gap} - In a few cases, the combination of a data parameter and an activity or location directly caused the data-expectation gap. For example, areas densely covered by buildings or trees often caused GPS issues. Participants also reported step underestimation \textit{because} they engaged in activities requiring holding or pushing, such as walking with a stroller. Here, the \textit{activity type} also led to an under-representation of effort as opposed to cases where steps were unrealistically overestimated.

\textbf{Temporality of Experiences with the Data-Expectation Gap} - Here, we borrow the term \textit{temporality} from the TaE framework~\cite{mccarthy2007technology}, which emphasises the temporal qualities of experience. In our studied sample, temporal aspects affecting the perception of the data-expectation gap included the \textit{temporal unit of analysis}, the \textit{moment of mismatch detection}, and the \textit{inconsistency in repeated experiences}.
\begin{itemize}
    \item \textbf{Temporal unit of analysis} - Participants based their evaluations on using temporal ``windows'' with varying sizes. The assessment was as a \textit{snapshot} (18\%), \textit{continuous} (15\%), or \textit{as a whole} (32\%). Sometimes, the same user used different units in different circumstances. 
    \begin{itemize}
        \item \textit{Snapshot} assessment - Participants often judged an entire activity based on a snapshot. For instance, they dismissed a night's sleep data because the device failed to record a brief bathroom visit. Consequently, users sometimes disregarded data that might have been accurate aside from a minor error.
        \item \textit{Continuous} assessment - Here, participants constantly checked the data during the activity it concerned, for feedback during exercise or to assess its quality by comparing it with another measure (\textit{\TDataMeasure}). This happened typically with HR or step data. 

        \item Assessment \textit{as a whole} - Others assessed whether the overall trends in the data aligned with their experience as a whole, e.g. noticing an unrealistically high floor count after a long day of work, without focusing on minute details.
    \end{itemize}
 
    \item \textbf{Moment of mismatch detection} - These differences in the temporal unit of analysis affected the \textit{moment of mismatch detection}, highlighting how users’ experiences vary depending on when they notice the data-expectation gap relative to the activity. Participants mentioned detecting a mismatch \textit{during} an activity (16\%), or \textit{after} it (13\%). 
    \begin{itemize}
        \item Detection \textit{during} an activity - Detecting a mismatch \textit{during} an activity or an emotional state it concerns, can hinder real-time feedback, disrupt immediate plans, or change the user's emotional state. Such instances were mostly connected to the context of exercise; participants, for example, noted inaccurate HR data during training, which hindered their plans to exercise according to HR zones using real-time feedback. GPS issues \textit{during} hiking also led to immediate pathfinding concerns for participants. Reviewing the data \textit{during} activities amplifying implausibility can also contribute to the perception of a \textit{\TDataLogic}. This effect is also related to the \textit{duration of the activity}; mismatches like detecting steps while on a couch or in bed might be difficult to notice if the time spent there is short, as the user may not be able to recall their exact behaviours. A \textit{\TDataMeasure}~involving comparison with training equipment can also only be detected \textit{during} exercise, as this is where the second data source is viewed for comparison. 
        \item Detection \textit{after} an activity - Identifying the data-expectation gap \textit{after} the activity primarily breeds mistrust in the data or affects the user's ability to understand their performance, impacting future reflections and plans, but without real-time consequences. Here, participants typically used a second measure as evidence (\textit{\TDataMeasure}), or evaluated the whole experience if this was easy - e.g. anticipating no climbed floors after sleep. 

    \end{itemize}

    \item \textbf{Inconsistency in repeated experiences} - We found few (2\%) instances where participants perceived the data-expectation gap \textit{through} repeated experiences. They doubted the data after noticing score variations in similar situations or identical data across different experiences; for example, recording different distances after doing the same route. Although very infrequent, this theme, which we call \textit{inconsistency in repeated experiences}, shows how repetition of experiences can also trigger disbelief. It implies a repeated interaction with the data, potentially dismissing a single event, but reflecting \textit{on}~\cite{mccarthy2007technology} multiple interactions and concluding that the data are unreliable.

\end{itemize}

\textbf{Co-experience} - Here we discuss how \textit{coexperience} affects perceptions of the data-expectation gap. While mismatches were mostly experienced individually, the social dimension was influential in a few cases (6\%). Similar to \textit{activity}, the presence of another individual \textit{caused or amplified disbelief}, or \textit{caused the data-expectation gap}. The other individual typically had another device, serving as a frame of reference. For example, when exercising or working with a partner, participants discovered mismatches by comparing devices, such as realising that their hiking device recorded a shorter distance than their friend's. In some cases, the other individual was a person with a higher authority relevant to the data - such as a trainer or doctor - who invalidated the data based on their knowledge or measurement with accurate equipment. While in most examples the second device helped identify the mismatch (\textit{\TDataMeasure}), in one case, two devices were purchased by a participant to track activities with a partner; both devices overestimated their activity, reinforcing the belief that activity tracking was unreliable in this model. The social dimension here was a factor \textit{amplifying disbelief}. Caring for a baby was a case where the presence of another individual indirectly \textit{caused the data-expectation gap}, as users move their hands a lot rocking the baby - which is mistaken as steps by the device - or push the stroller, which leads to step underestimation. Notably, one participant reported a malfunctioning fall detector that misclassified breaks for drinking water as falls, resulting in unnecessary calls to their husband or 911. This case shows how the data-expectation gap can also affect others.

\textbf{Personal factors} - Personal factors influenced participants' perceptions of the data-expectation gap; namely, their \textbf{goals and motivations} and \textbf{background}. Participants who were less concerned with a specific metric, either because it was not central to their interests or because the device met their primary needs, tended to overlook mismatches (14\%). In contrast, mismatches that created barriers to the participants' goals, caused frustration (13\%). These goals and motivations were diverse; some (6\%) specifically desired accuracy during training, while others bought it for increasing motivation, general activity tracking, reliable HR monitoring, tracking heart palpitations, or travelling safely. For example, mismatches between actual and detected activity rendered useless the feature of challenges for a participant who had bought the device for this feature. A few participants (4\%) also reported health-related issues which affected the acceptance of mismatches; for example, one bought the device after heart surgery, noting that they were bothered by wrong HR measurements, while accurate ECG results were important for another participant in a long wait for a formal ECG test.

\textbf{History of experiences} - Another factor changing the significance of a mismatch is the history of encounters with other mismatches. Understanding this history is key to understanding how mistrust is developed; mistrust can increase through the accumulation of related experiences concerning the same or different parameters, or it can develop after a single experience and concern one data type or multiple. These different mechanisms are related to the following subthemes: the \textit{frequency of encounters}, the \textit{co-occurrence of mismatches}, the \textit{cascading effect} of a mismatch, and \textit{prior encounters} with the data-expectation gap. 
\begin{itemize}
    \item \textbf{Frequency of encounters} with the data-expectation gap - Many participants (50\%) mentioned repeated experiences with the data-expectation gap, describing related behaviours as something the device usually does. In the other half of the analysed reviews, the frequency of encounters with the data-expectation gap was unclear. Few participants, though (4\%) perceived only one or two mismatches, which were enough for them to conclude that the device was useless. In these cases, participants had a heightened attention for mismatches, with some describing their first interaction with the data as a purposeful ``test'' the device failed; such as a test run where the recorded path or distance deviated from reality. 
    \item \textbf{Co-occurrence of mismatches} - Participants also sometimes identified multiple types of mismatches. These issues were identified \textit{between-parameter} or \textit{within-parameter}, with both types potentially increasing the perceived unreliability of the data.
    \begin{itemize}
        \item \textit{Between-parameter co-occurrence of mismatches} (23\%) include cases where participants encountered mismatches in two or more different parameters. These cases acted as evidence that multiple data types are unreliable, amplifying mistrust. Some participants also reported more than two types of mismatches, and in different conditions; one participant, for instance, mentioned an underestimation of their speed while running in the forest, HR values that did not align with activity intensity during hiking, and consistent underestimation of climbed floors and fluctuating altitude at home. 

        \item \textit{Within-parameter co-occurrence of mismatches} (6\%) refer to mismatches within the same parameter, under different or opposing conditions. For example, some found their HR lower than expected during exercise, and higher than expected during rest; similarly, a participant observed a large number of steps detected while using a tractor, but much fewer than expected steps detected while on the treadmill and the elliptical. These combinations generate the impression that nothing works in this parameter.

    \end{itemize}
    
    \item \textbf{Cascading effect} of a mismatch - This category describes a few cases (5\%) where the detection of a mismatch actively triggered other mismatches, or caused quality concerns for other data types. Such mismatches led to negative sentiments and caused participants to not recommend the device. The \textit{cascading effect} may be \textit{parameter-specific}, or \textit{generalised}. The \textit{parameter-specific} subtype leads to further mismatches in derivatives of the first parameter. For example, GPS problems caused problems in speed detection during exercise, which then led to delayed and irrelevant pace alerts. Another participant's smartwatch consistently underestimated their HR; on a few occasions where the participant perceived the HR as reasonable, the watch perceived it as above average, consequently underestimating their readiness score. Other users may develop mistrust for parameters related to the mismatch, without actually encountering mismatches in these parameters; for example, a mismatch between HR data on the smartwatch and another device during exercise rendered all data related to HR untrustworthy for a participant, including calories, minutes in HR zones, and resting HR. In the \textit{generalised} subtype, this mistrust extends beyond related parameters to affect the credibility of all data types. One participant, for example, expressed that since GPS was erroneous, they assumed all other features, such as workout tracking, HR monitoring, and sleep data, would also be unreliable, despite no direct link between GPS and these features.

    \item \textbf{Prior encounters} with the data-expectation gap - This category refers to interactions with data from other devices, before or alongside using the current device. Prior encounters with mismatches shape expectations about data trustworthiness, and each new experience influences future perceptions. 40\% of participants had previous experience with another activity tracking device, 3\% had no experience and for the rest, it was unclear. Previous devices acted as a frame of reference for many participants. Some participants, based on past experiences with similar devices, anticipated certain mismatches, or noted that the new device had some shortcomings in data quality but was performing better than their previous device. One participant, for example, noted that their new device's reported HR differed from their chest strap - which they trusted more - during sports, but overall HR monitoring was better than their previous model. Others were disappointed when heightened expectations were not met - particularly when they already had a previous device which was reliably tracking the same data (13\%), or when the mismatch concerned pricier models which they bought hoping for high accuracy (7\%). Some of those participants with high expectations due to price were particularly frustrated when they perceived a detected mismatch as a failure of the watch to perform a core function, such as tracking steps, HR or GPS (4\%).

\end{itemize}

\subsection{Key findings}

The first study resulted in the following key themes related to the data-expectation gap:  

\begin{itemize}
    \item \textit{Source of evidence} - Users identify mismatches using different sources of evidence, which affect the degree of implausibility of the data.
    \item \textit{Types of mismatches} - The identified mismatches have various types regarding the identification of a wrong state or value.
    \item Contextual factors - \textit{activity}, \textit{location}, \textit{temporality} and \textit{co-experience} affect considerably the implications of each type of mismatch.
    \item \textit{Goals, motivations and background} - The perception of a mismatch is also influenced by personal factors.
    \item \textit{History of experiences} - The history of encounters with various mismatches influences how a new mismatch is perceived. Prior experiences with data from other devices shape expectations. Mistrust can grow through the accumulation of related experiences, involving the same or different parameters, or it can arise from a single experience.
\end{itemize}

%% file: Sections/4b-FieldStudyResults.tex
\section{Study 2 - In-the-Wild Experiences of the Data-Expectation Gap}

Whereas the first study revealed factors influencing perceptions of the data-expectation gap, online reviews reveal only curated snapshots of users' experiences. Therefore, we decided to complement the first study with a second study involving a more in-depth data collection from fewer participants, aiming to enrich our insights and deepen our understanding of participants' experiences with the data-expectation gap and its intersections with everyday life. We conducted a three-week field study involving a smartwatch and an experience sampling app. We interviewed the participants, and analysed their recollections and the field data, aiming again to identify factors affecting how they perceived their encounters with the data-expectation gap.

\subsection{Study Design}

\begin{table*}[h]
    \centering
    \includegraphics[width=1\textwidth]{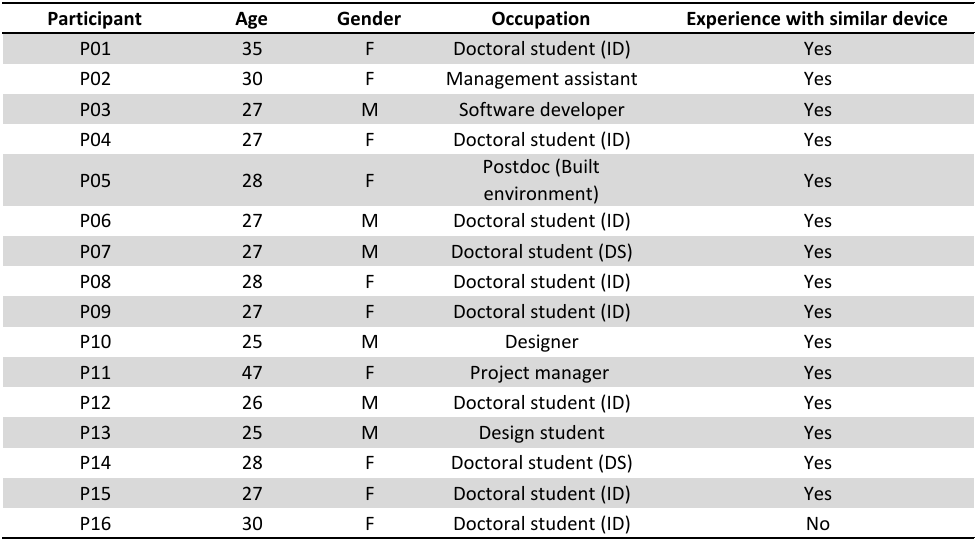}
    \caption{Participant characteristics. ID: Industrial Design. DS: Data Science}
    \label{table:Participants}
\end{table*}

\subsubsection{Participants}
We recruited 16 participants (median age: 27, range: 25-47), including ten females and six males (Table~\ref{table:Participants}). We used convenience sampling, as the study required three weeks of equipment use without compensation. Most participants (n = 11) came from the local university. Data collection spanned November 2022 to March 2023 and the study had three phases:
\begin{itemize}
    \item initial meeting for informed consent and on-boarding,
    \item three-week field experiment,
    \item post-study interview.
\end{itemize}

\subsubsection{Apparatus}

\begin{figure*}[ht]
\centering
\begin{minipage}{0.34\textwidth}
  \centering

  \includegraphics[height=5cm]{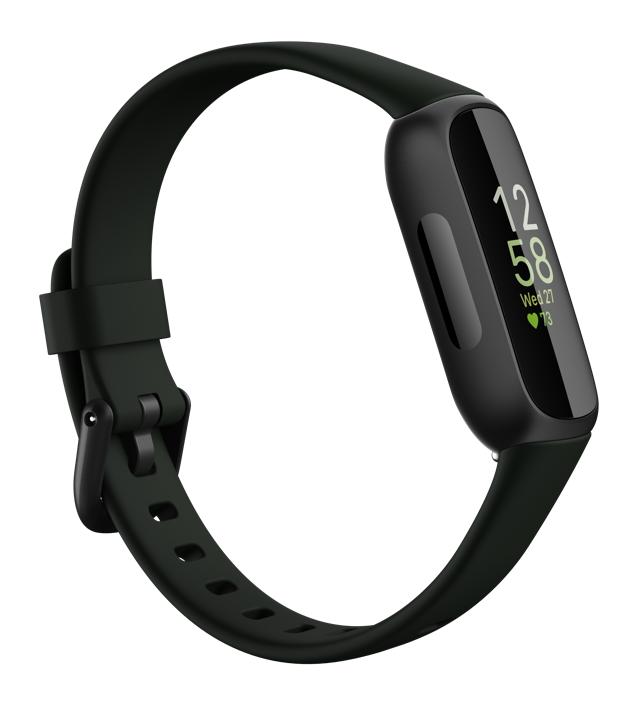}
  \captionof{figure}{Fitbit Inspire 3, which was used in the field study (Figure retrieved by Fitbit~\cite{Fitbit2023}). }
  \label{fig:inspire3}
\end{minipage}%
\vspace{-1.5em}
\begin{minipage}{0.65\textwidth}
    \centering
    \captionsetup{width=0.9\linewidth}
    \includegraphics[height=5cm]{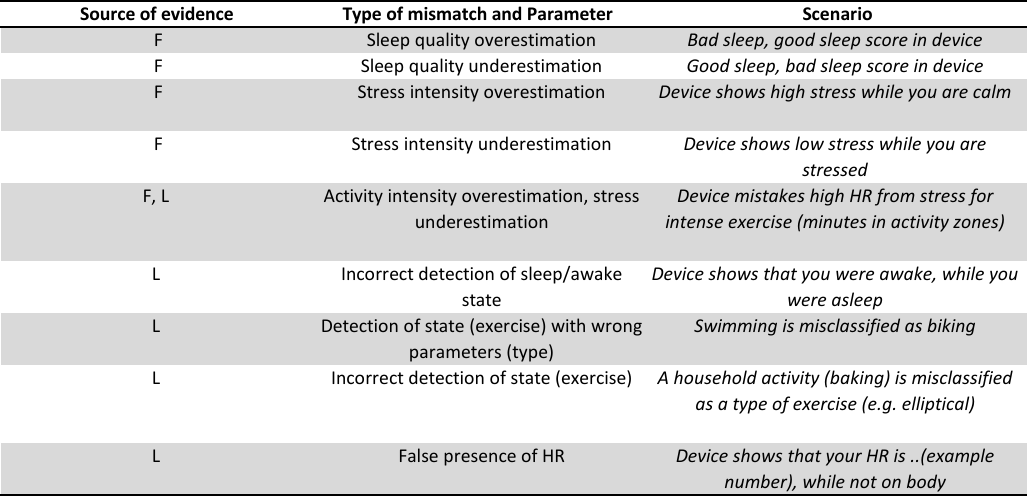}
    \caption{The discussed scenarios of mismatches. F: \TDataFeeling. L: \TDataLogic.}
    \label{table:scenarios}
\end{minipage}%
\end{figure*}

Participants received a Fitbit Inspire 3 (Figure \ref{fig:inspire3}) for three weeks and were free to use it as desired. This device was chosen based on local availability, budget limitations and widespread brand use~\cite{Fitbitpopularity2024}, as an example of an affordable smartwatch measuring various data types that could induce different experiences. This device supports the same metrics as Fitbit Sense 2 from our first study, excluding some stress-related features: EDA and continuous EDA (cEDA) tracking, stress notifications and relaxation prompts. Further details are provided in \ref{Appendix: Fitbit characteristics}.

Additionally, we asked participants to use a smartphone application (m-Path~\cite{mestdagh2022mPath}) for gathering real-time information on experiences using the Experience Sampling Method (ESM)~\cite{hektner2007ESM}. We chose ESM over a diary study, as we wanted to collect contextual information with high frequency and avoid cases where participants might be overly selective in their diary entries~\cite{rapp2016personal}, discarding parameters that could be of interest to us. Each ESM questionnaire took about one minute, could be skipped, and expired after a set time. The ESM protocol, finalised after a three-week pilot with one participant, involved three questionnaires with 7-point Likert items, multiple-choice questions, and open-text fields. The themes per questionnaire were as follows:
\begin{itemize}
    \item \textbf{Morning questionnaire} - SQ, readiness, work mode, stress
    \item \textbf{Four-hourly questionnaire} - Location, company, activity, mood, fatigue, stress, interesting events, type of conducted exercise
    \item \textbf{Evening questionnaire} - Exercise obstacles, issues with equipment.
\end{itemize}
The ESM data that could be directly related to Fitbit data were subjective stress, subjective SQ and logged activities. The other ESM questions aimed to provide context. An open-text field allowed recording activities not included in our list. More details are provided in \ref{Appendix: ESM protocol}.

\subsubsection{Interview Protocol}

Each participant was interviewed shortly after their field data collection. These semi-structured interviews, conducted in English, lasted 60-90 minutes. All participants and the interviewer were non-native English speakers. To understand participants' experiences with the data-expectation gap, we discussed scenarios (Figure~\ref{table:scenarios}) like receiving a poor sleep score despite feeling rested. Participants shared experiences concerning each scenario, including experiences before the field experiment if relevant. To benchmark participants' ESM responses~\cite{Cury2019EthnographyML}, we also discussed instances where participants used extreme Likert item scores. While the interviews also covered other aspects like opinions on the ESM app (see protocol in \ref{Appendix: Interview protocol}), our analysis focused on the data-expectation gap. All sessions were audio-recorded.

\subsubsection{Data Analysis} \label{Field study:data analysis}
We analysed the interview data using reflexive thematic analysis~\cite{Braun2019ReflexiveTA}. The lead author conducted the interviews, transcribed the data verbatim and coded the data using NVivo 14. Initial codes were generated using inductive analysis, followed by deductive analysis influenced by the same themes as the deductive analysis of Study 1, and the findings of Study 1. The initial set of 49 codes was grouped into three themes (\textit{emotional state}, \textit{trust}, \textit{system intelligibility and participants' mental models}), finalised after deliberation with the second author.

The analysed field data comprised Fitbit data (resting HR, hourly mean HR, hourly step count, sleep score, stress management score, and detected or logged activities) and ESM data. We identified potential mismatches by comparing sensor and ESM data (details are provided in \ref{Appendix: Fitbit-ESM alignment}) for: subjective stress versus Fitbit stress management score, subjective SQ and Fitbit sleep score, and logged versus Fitbit-detected exercises. We analysed how many mismatches participants had the potential to experience based on data logs, tracked participants' mentions of related instances, and studied the context of these experiences based on the ESM data.

\subsection{Results}

\subsubsection{Field Data Insights}

\begin{table*}[h]
    \centering
    \includegraphics[width=0.8\textwidth]{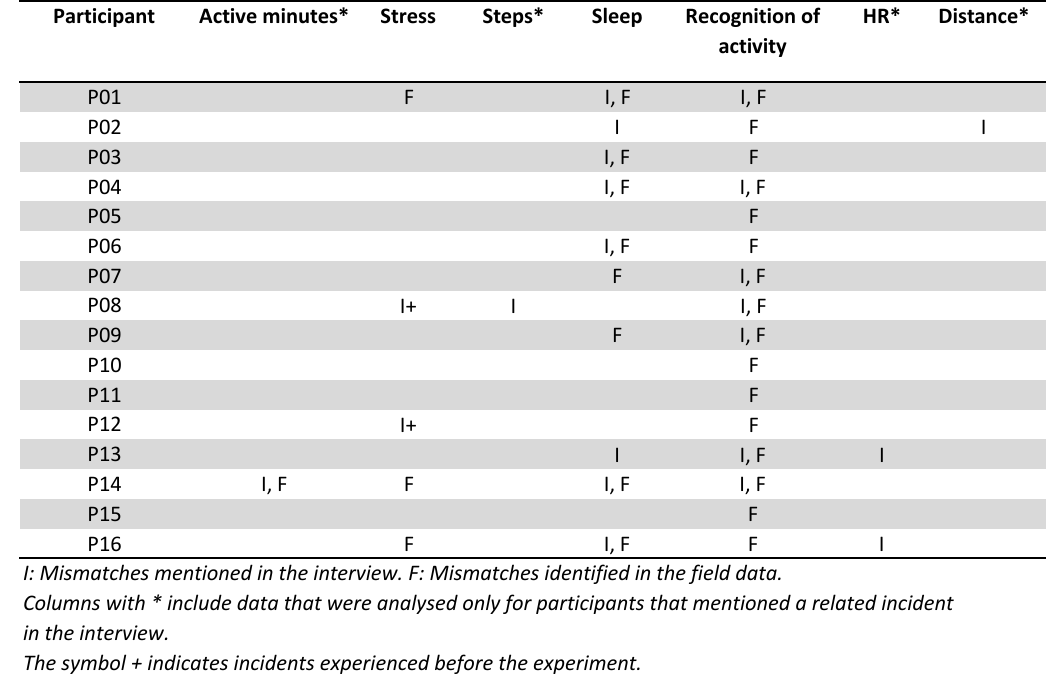}
    \caption{Mismatches mentioned in interviews or identified in the field data.}
    \label{fig:inaccuracies_per_user}
\end{table*}

Field data corroborated most recollections (Table~\ref{fig:inaccuracies_per_user}). Our main finding from this analysis was the notion of \textit{tension potential}, connected to \textit{interaction patterns}, \textit{tendency to identify activities of a specific type}, \textit{varying success in correct classification of activities between and within participants}, \textit{differences in the contextual character of mismatches versus successful detections}, and \textit{sequentiality of mismatches}.

\textbf{Tension potential} - While field data showed a high potential for experiencing mismatches in activity classification and SQ estimation, the mismatches mentioned in interviews were notably fewer. For example, while Fitbit failed to classify a workout and a sports activity of P03, and misclassified an activity within its capabilities, P03 only mentioned a successful classification. The low recall rate for activity classification might stem from not checking activities in the app. We introduce the term \textit{tension potential} to describe that while smartwatches can create multiple possibilities for encountering mismatches, participants may perceive only a few depending on their interests, interactions with the device, and context. We also define \textit{potential mismatches} as possible mismatches based solely on the data, when it is unclear whether participants would perceive each instance as a mismatch.

\begin{figure*}
\centering
\begin{minipage}{1\textwidth}
  \centering
  \captionsetup{width=1\linewidth}
  \includegraphics[width=\textwidth]{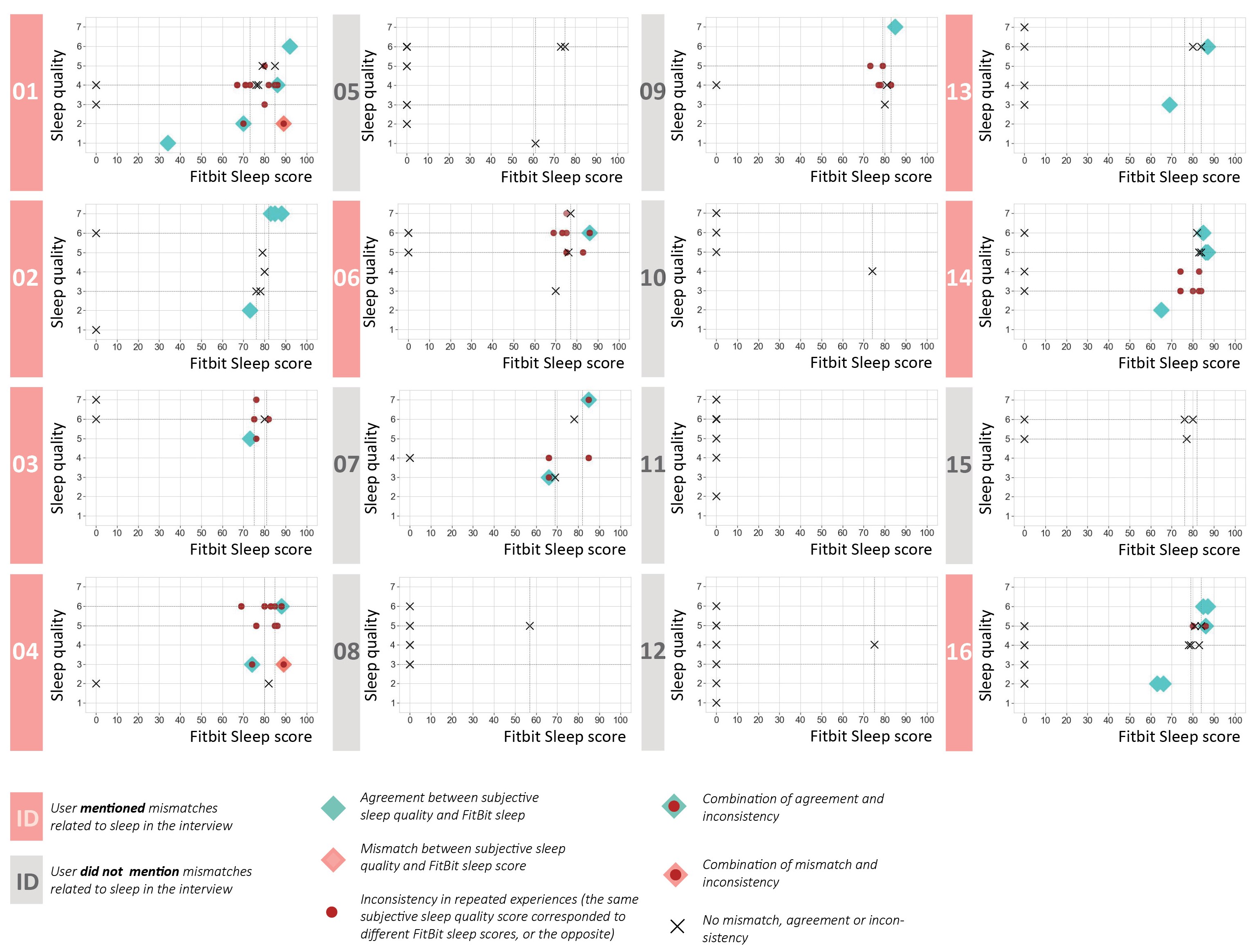}
  \captionof{figure}{Evidence of the data-expectation gap in SQ.}
  \label{fig:disagreements_sleep}
\end{minipage}%
\vspace{-1.5em}
\end{figure*}

\begin{figure*}
\centering
\begin{minipage}{1\textwidth}
  \centering
  \captionsetup{width=1\linewidth}
  \includegraphics[width=0.87\textwidth]{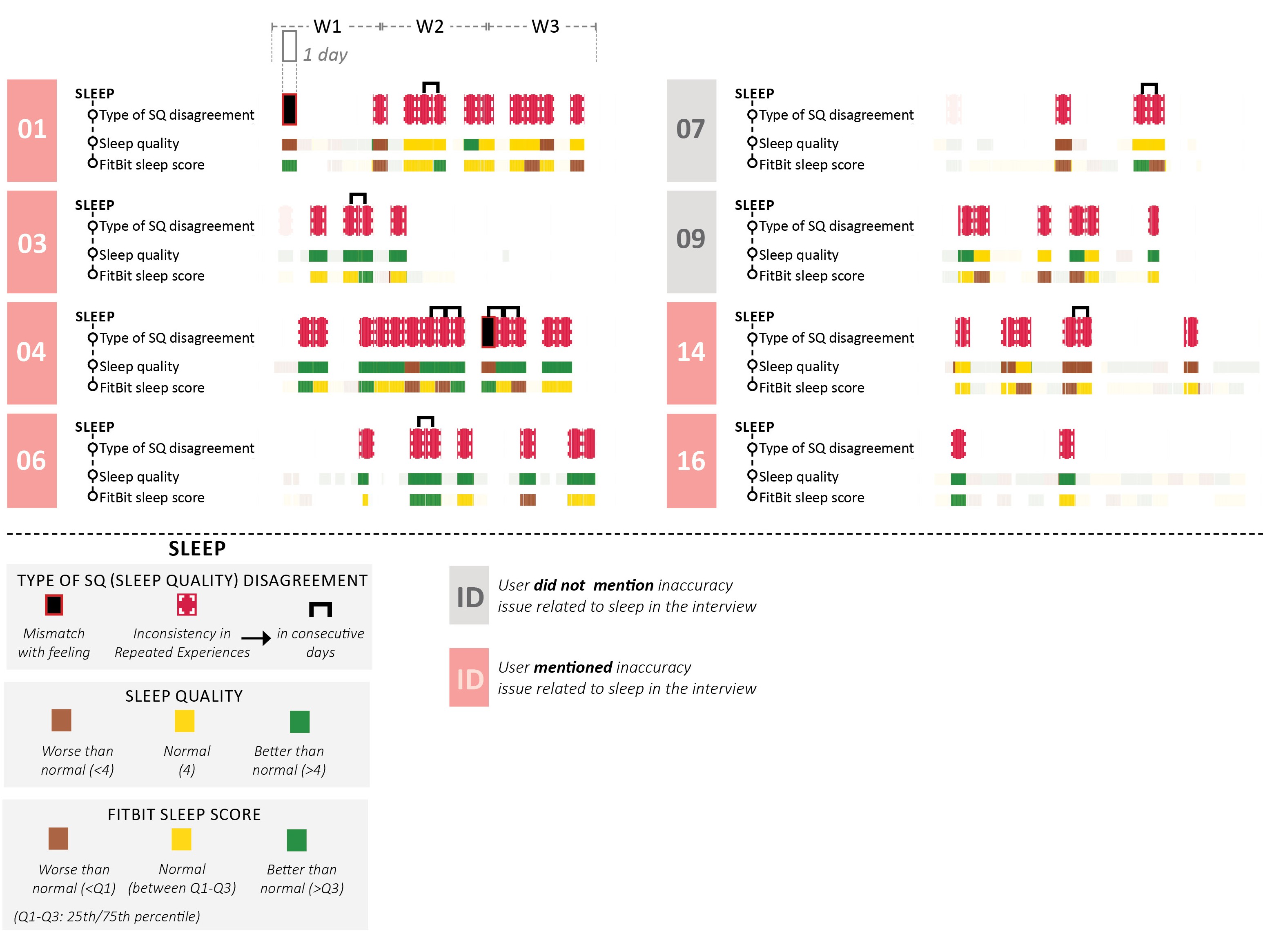}
  \captionof{figure}{Potential mismatches between perceived and detected SQ during the field experiment, and related contextual data. The data are visualised as a timeline for each participant, with each color bar representing a unit of 4 hours. \textit{Potential mismatches} are visualised in the first row of each participant's data.}
  \label{fig:sleep_disagreements_details}
\end{minipage}%
\vspace{-1.5em}
\end{figure*}

Figure \ref{fig:disagreements_sleep} exemplifies differences in \textit{tension potential} among participants, based on our SQ data analysis. It shows two potential instances of a \textit{\TDataFeeling} -  marked with light pink, diamond-shaped markers - where Fitbit overestimated sleep scores for P01 and P04, which they also recalled. We also found possible \textit{inconsistencies in repeated experiences}, shown with crimson round markers, in the sleep data of eight participants. For example, on two occasions, P07's Fitbit sleep score was 85, but they self-rated it with 4 and 7. However, only P03 and P14 recalled such experiences. Therefore, while these eight participants had high \textit{tension potential}, it only materialised for P03 and P14. Figure \ref{fig:sleep_disagreements_details} shows that P01's and P04's SQ field data indicate more \textit{potential mismatches} than others,  meaning they had the highest \textit{tension potential} and a higher chance of experiencing a mismatch.

\begin{table*}[h]
    \centering
    \includegraphics[width=1\textwidth]{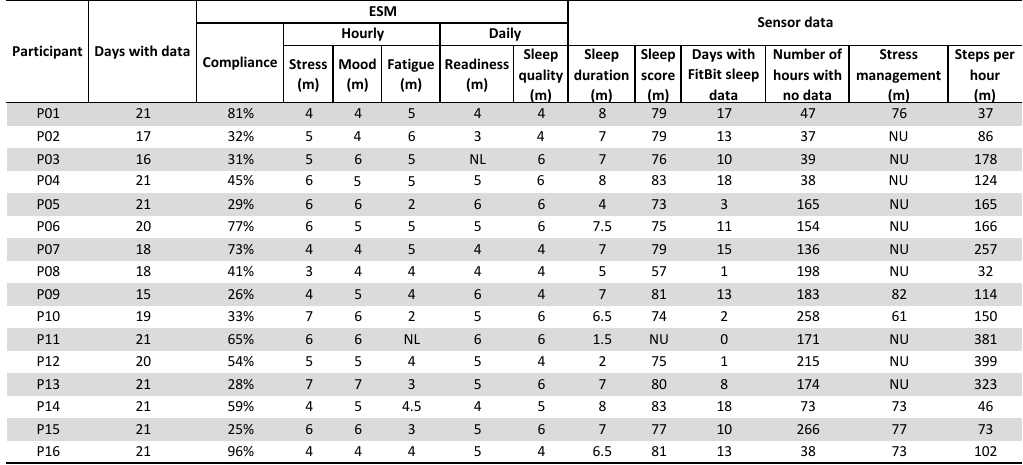}
    \caption{Overview of field data. NL: No value logged. NU: Feature not used. Row names including (m) show median values, calculated using non-zero values (apart from step data).}
    \label{tab:field-data-overview}
\end{table*}

\begin{figure*}[h]
    \centering
    \includegraphics[width=0.7\textwidth]{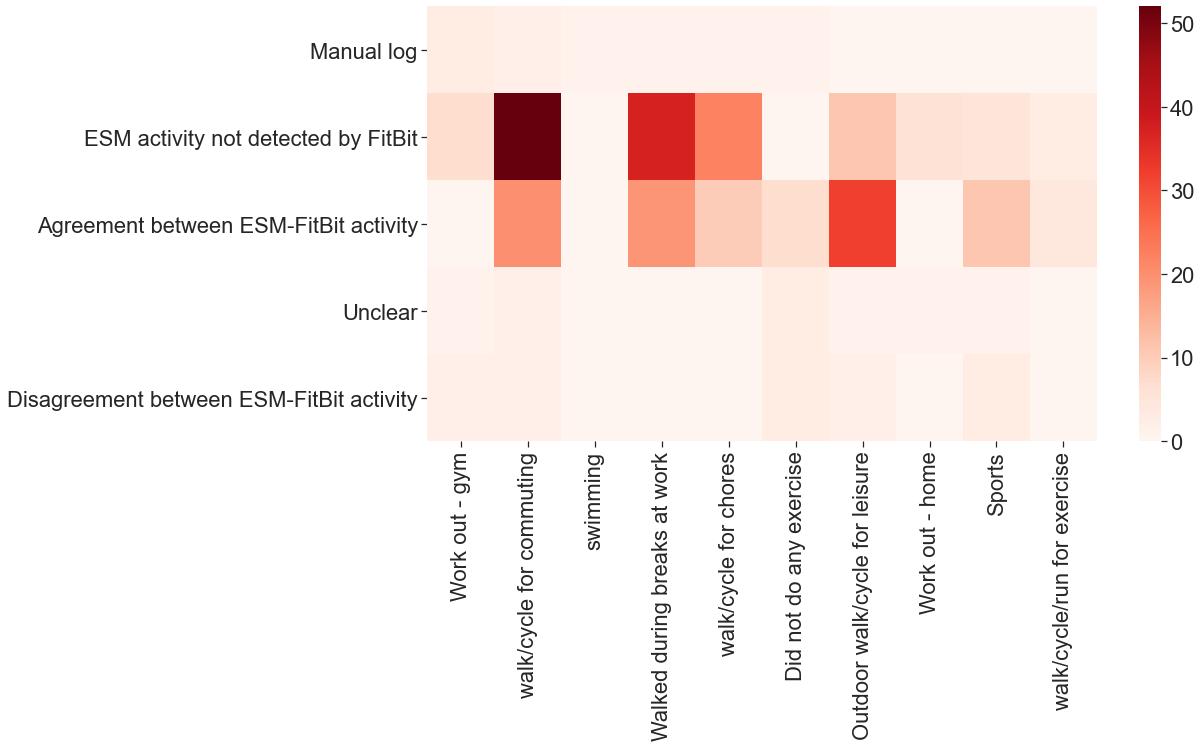}
    \caption{Comparative analysis of activities in ESM and Fitbit logs.}
    \label{table:activ_agreement}
\end{figure*}

\textbf{Interaction patterns} - Table~\ref{tab:field-data-overview} shows that device usage patterns varied during the study. The \textit{tension potential} is related to these variations. Few used the Fitbit stress management feature, and some primarily wore it while awake. A few (n=3) even mentioned never checking the Fitbit app, due to indifference or preference for another smartwatch. Therefore, any potential mismatches in these participants' data did not materialise due to not interacting with the data - because of disinterest in sleep tracking, disinterest in the app, or complex user interface (UI), based on the interviews.

\textbf{Tendency to identify specific activities} - Fitbit often recognised leisure walks or cycles logged in the ESM app but typically missed commuting activities (Figure~\ref{table:activ_agreement}). Therefore, participants doing more commuting activities had higher \textit{tension potential}. We found six instances of activity misclassification, and a case where the watch detected activity during rest. Some of these \textit{potential mismatches} were recalled in interviews, others were not (Table~\ref{fig:inaccuracies_per_user}).

\begin{figure*}
\centering
\begin{minipage}{1\textwidth}
  \centering
  \captionsetup{width=1\linewidth}
  \includegraphics[width=0.9\textwidth]{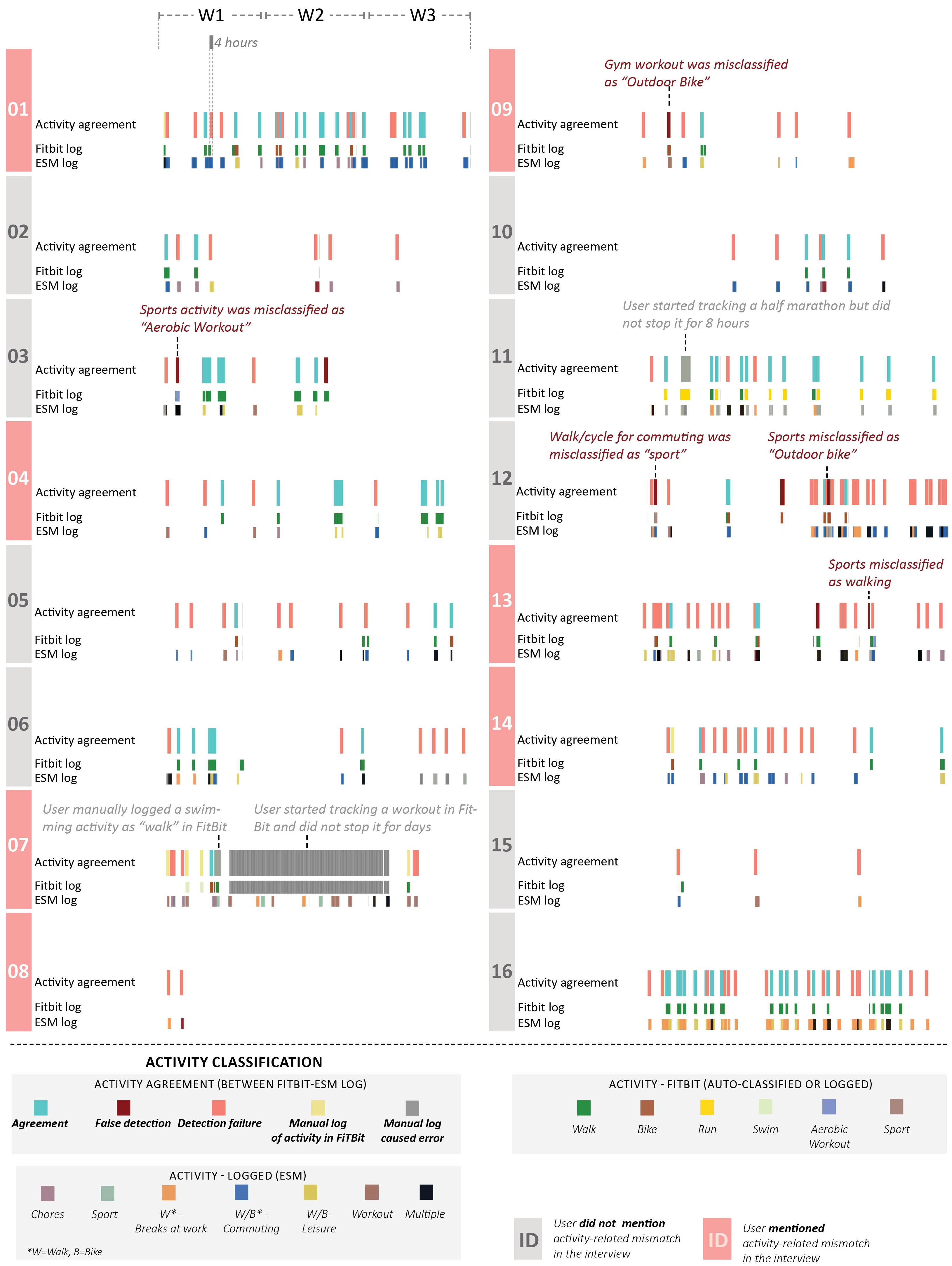}
  \captionof{figure}{Alignment between ESM and Fitbit activity logs, and related contextual data. The data are visualised as a timeline for each participant, with each color bar representing a unit of 4 hours. \textit{Potential mismatches} are visualised in the first row of each participant's data, with light pink (indicating \textit{detection failure}) and crimson (indicating \textit{false detection}). Cases where the device output agrees with participants' logs are marked with cyan.}
  \label{fig:activity_disagreements}
\end{minipage}%
\vspace{-1.5em}
\end{figure*}

\begin{figure*}
\centering
\begin{minipage}{1\textwidth}
  \centering
  \captionsetup{width=1\linewidth}
  \includegraphics[width=\textwidth]{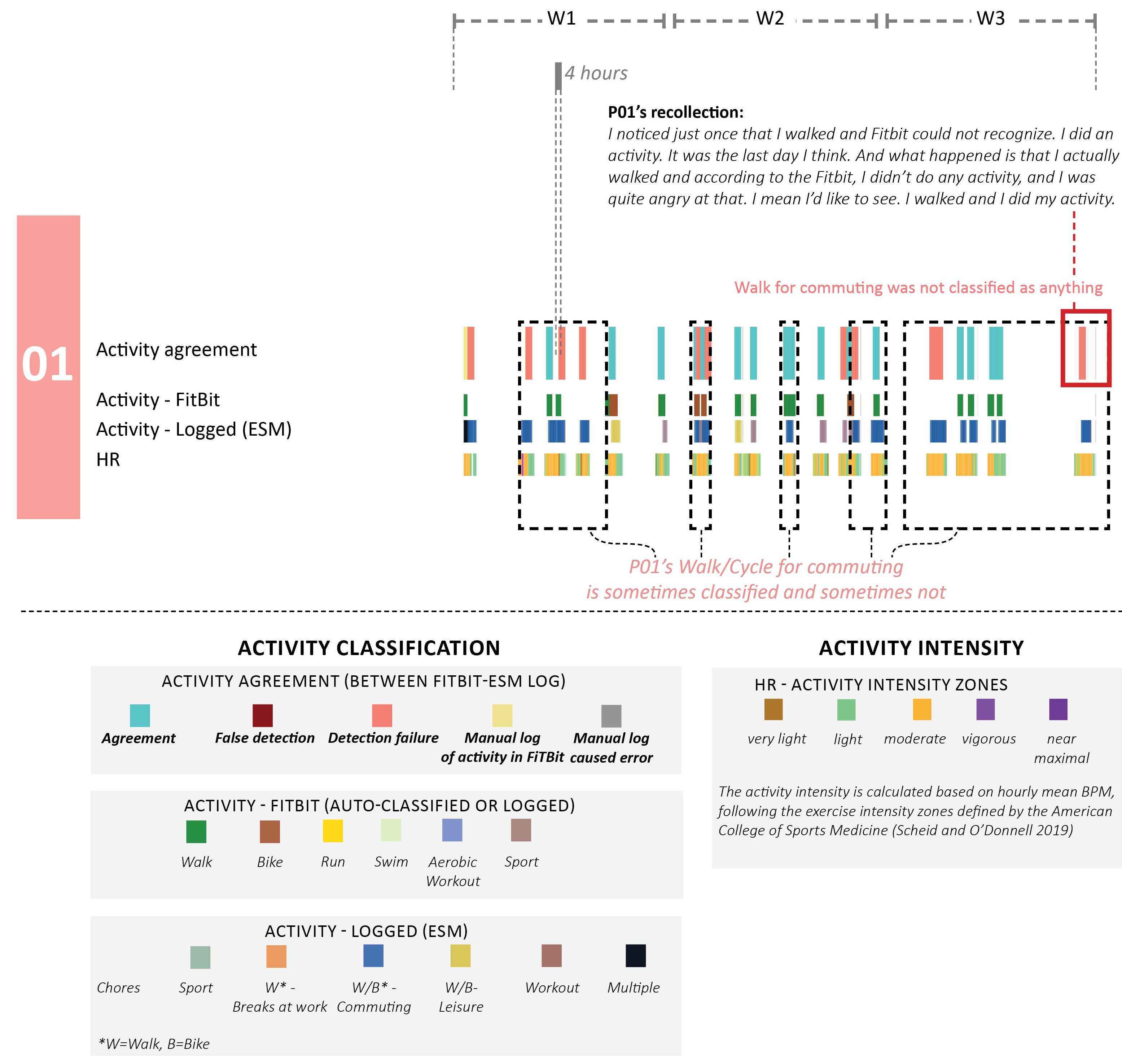}
  \captionof{figure}{Alignment between P01's ESM and Fitbit activity logs, represented following the same scheme as Figure~\ref{fig:activity_disagreements}.}
  \label{fig:P01Activity}
\end{minipage}%
\vspace{-1.5em}
\end{figure*}

\textbf{Varying success in correct classification of activities} - Figure~\ref{fig:activity_disagreements} shows the \textit{potential mismatches} in activity data per user, illustrating that some participants had more successfully recognised than failed activities, and some had the opposite. Certain activities were recognised for some, but not for others. For example, walks during breaks at work were recognised as an activity for P11, but never for P09. The classification success also sometimes varied within participants for the same activity. Figure~\ref{fig:P01Activity} details P01's activity data, showing 15 successful activity classifications and 10 \textit{detection failures}. Most logged activities were walks for commuting, which Fitbit correctly identified nine times and did not identify seven times. P01 only noted one \textit{detection failure} for this activity, but noted their frustration with this failure. Therefore, the \textit{tension potential} only materialised in this instance. The figure also displays their HR data which were in the moderate intensity zone, confirming their perception of this activity as intense.

\textbf{Contextual characteristics of mismatches and successful detections} - Participants' mood was more frequently positive in activities Fitbit did not recognise (77\% vs. 63\% in successfully detected activities), and activity intensity was lower (HR was 68\% in moderate or higher zones, vs. 92\%). Social presence was similar, with company 50\% of the time. To better understand these differences, we further analysed the types of activities users logged. Fitbit labelled as `walks' diverse and frequently social activities which participants might not necessarily want to track - from pub-crawling (P03) to meeting a friend with their baby (P14) - but often missed workouts. For example, it recognised P04's leisure walk to a local festival but did not recognise their workout, which they wanted to track. (P04) \textit{``It only recognised walks...I had a couple of workouts where I really pushed myself...And I was really excited. Then I saw that there's nothing.''} P04's account also highlights the intensity of some unrecognised activities, which is related to the \textit{sensual} thread of experience~\cite{mccarthy2007technology}. These mixed success rates might influence users' mental models of the device's capabilities.

\textbf{Sequentiality of mismatches} - Figure \ref{fig:sleep_disagreements_details} also shows another property in the \textit{temporality} of the data-expectation gap: \textit{sequentiality}, where the \textit{\TRepeatedMeasures} concerns two consecutive days, making it easier to compare them and perceive a mismatch. The figure also shows that P01, P04, P07 and P14 sometimes had bad SQ on days with potential mismatches, while the others had slept normally or well - a difference which may have affected participants' perception of mismatches.

\subsubsection{Interview Insights}

Most interviewees (n = 13) recounted experiences with the data-expectation gap in smartwatches, during or before the experiment. The ambiguity of the Fitbit sleep score was highlighted, as some interpreted it by comparing it to previous scores, while others considered the watch descriptors, such as ``Good'' or ``Fair''. Although accuracy was desired, a few also experienced tensions when data confirmed an undesired state. (P01) \textit{``...I felt this sense of oppression, and also a little bit of heart beat. I used the Fitbit to see what happened. And I noticed that actually I had a slightly faster HR... So again you have a sort of confirmation that there is something wrong.''}

\textbf{Identification of themes from the previous study} - We first discuss the applicability of themes identified in the previous study: \textit{source of evidence}, \textit{types of mismatches}, contextual factors, users' \textit{goals, motivations and background}, and \textit{history of experiences}. Regarding the \textit{source of evidence}, the \textit{\TDataLogic} theme was the most prominent (n=11), as in the online reviews. The \textit{\TDataFeeling} theme appeared more frequently (n=7) compared to the online reviews. (P03) \textit{``Even in the night that I had insomnia and I was waking up and I was feeling like trash I got 86.''} Fewer participants (n=3) mentioned the \textit{\TDataMeasure} theme. Regarding \textit{types of mismatches}, the most frequently mentioned issues involved \textit{overestimation} (n=11) concerning value estimation, and \textit{false detection} (n=9) concerning classification. These concerned various parameters - SQ, steps, stress, HR, and activity.

As in our previous study, contextual factors affected the perception of these mismatch types. P13's recount, for example, shows how the \textit{location} and \textit{activity} during a mismatch caused worry, while the \textit{co-experience} eased tension: \textit{``When I looked back at my data that sometimes my HR at the beginning of the night was very heavy and I was just sitting or I was just looking at TV. So eventually I was checking, should I go to a doctor because I have random peaks? Luckily, my girlfriend is studying medicine...And she was like, `No, it's fairly normal.' ''} In P02's recollection of a \textit{\TDataLogic}, \textit{activity intensity} was the factor assessed to identify the mismatch, and the \textit{co-experience} again eased tension: \textit{``I went to Zumba with P07 and then it said that I walked 500 kilometers. And I think, no shit, that's not true.[] We both had a look and started laughing.''} Two participants reported \textit{inconsistency in repeated experiences}, which was connected to \textit{temporality} in our previous study. P16 mentioned, \textit{``I think the range on the score from the Fitbit is very small. Sometimes I feel `ohh I sleep, I sleep well' and I got maybe 80 points there, but sometimes I feel I didn't sleep very well, but I still got maybe 76 points.''} We also identified a new contextual factor influencing the perception of mismatches: \textit{emotional state}. We later discuss this theme in detail.

Users' \textit{goals, motivations and background} were again influential. Similar to the online reviews, participants had stronger opinions on mismatches concerning data relevant to their interests (n=5). Mismatches leading to a misrepresentation of their effort related to an interest created frustration. P01 noted, \textit{``I walked and I did my activity. I did that but it was not there, it didn’t record it. And so that's annoying.''} P09 similarly remarked, \textit{``I also do other activities like workouts and jumping, but I can't find them. So I’d have a feeling like I have exercised a lot but the system doesn’t know it.''}

Regarding the \textit{history of experiences}, some participants (n=3) recalled cases where a mismatch was initially acceptable but caused mistrust after repetitions. This is related to the \textit{frequency of encounters} with the data-expectation gap. We also found \textit{co-occurrence of mismatches} (n=9); Some participants experienced mismatches in multiple parameters (\textit{co-occurrence of mismatches}), such as in stress and activity, while a few experienced different types of mismatches within the same parameter, such as overestimation and underestimation of SQ. (P03) \textit{``...the sleep is off, the stress is off, or it shows HR while not wearing. If I see a couple of them for me, there's like a bigger implementation problem. And I’ll take all the data with a pinch of salt.''} We also found the theme of \textit{prior encounters} with the data-expectation gap, which we later discuss in the context of our new findings.

The themes identified in the previous study were, therefore, also present in this study. From now onwards, we focus on the new parameters we identified as influential regarding the perception of mismatches: \textit{emotional state}, \textit{trust}, and \textit{explainability and mental models}.

\textbf{Emotional state} - The valence of the emotional state affected the perception of the data-expectation gap, when connected to a \textit{\TDataFeeling}. Seven participants mentioned having negative feelings - low SQ or high stress - which the device disconfirmed, while three mentioned a disconfirmed positive feeling. Conceding to the algorithm sometimes improved and sometimes worsened the emotional state, while some participants also maintained their state, because they did not find this a reason to override their positive feeling, or they trusted their feelings more in these areas.

\textit{Changing emotional state} - Two participants reported \textit{improving their emotional state}, accepting the unexpected results. They were initially feeling stressed, and calmed down after seeing the data indicating a better state. P13 recalled: \textit{``And I was checking it and it was just normal like my normal days. So I was like oh, then I'm just fucking myself up. My HR isn't that high.''} P3 similarly conceded to the algorithm when seeing that the detected SQ was unexpectedly high, as this overestimation of their state motivated them to start their day feeling more energised. 

Some participants also reported an experience of \textit{worsening their emotional state}. P13 initially felt satisfied with their sleep, but worried upon seeing data indicating a worse state: \textit{``But then by seeing it saying it wasn't that good, I felt like I didn't sleep that good. Then I was a bit demotivated for the day.''} Four participants felt worse when the device misclassified a negative state as positive - mostly due to frustration that their negative perception of SQ was not validated. These participants were very certain about their feelings, and received a good score while feeling tired and frustrated from insomnia or having to wake up early. P03 highlighted their heightened sensitivity due to their tiredness: \textit{``I think in that certain situation it [validation] would have helped me, because I was feeling so different from yesterday, and I carried a bit of frustration from trying to sleep when I woke up[...] And also Fitbit said `Ohh yeah you slept great. Let's go. Let's talk at the world.' It's like, `No, you don't know half the story.'''} Similarly, P06 was surprised by an overly positive sleep score: \textit{``It says now that I slept for I think 7+ hours and I slept well. And I was like, no, that wasn't the case.''} P06 had unsuccessfully tried to sleep early enough to get fully rested, and the overestimation of their detected sleep hours on top of the wrong SQ score made the data implausible, making this a combination of a~\textit{\TDataFeeling}~and a~\textit{\TDataLogic}. 

\textit{Maintaining emotional state} - Some participants also maintained their emotional state after a mismatch. P03 for example felt validation unnecessary when getting a bad SQ score while feeling good, unlike earlier frustration with an unexpectedly good score: \textit{``There's no change to my mood because if I feel refreshed, it doesn't matter that much.''} Three others also noted the mismatches but did not get particularly affected, ignoring them. For example, while P03, P04 and P016 all have sleep issues, P16 ignored sleep-related mismatches that had frustrated P03 and P04, possibly because of differences in \textit{prior encounters} and expectations; P03 and P04 had other smartwatches before, which performed better in sleep reporting. This was, though, P16's first experience with a tool that could help them understand their sleeping pattern, and valued it due to its perceived accuracy in other instances. Ignoring mismatches may also be a behaviour adopted over time, recalling the \textit{frequency of encounters} which we discussed in the previous study. P08 characteristically reported, \textit{``according to Garmin, I'm 100\% stressed...If I'm, for instance, biking home or walking faster while I'm actually taking breaks''}, a mismatch so frequent that by now they have learnt to ignore it.

\textbf{Trust} - Most of the themes discussed until now already suggest how encounters with mismatches lead to mistrust. Here, we further discuss how participants' trust \textit{in the self} or \textit{in technology} affected perceptions of the data-expectation gap.

\textit{Trust in technology} - Behaviours related to the~\textit{\TDataFeeling} were mediated by participants' trust towards technology and themselves. Some reported becoming uncertain about their feelings, trusting the algorithm more. P02 humorously commented on their reliance on technology: \textit{``...you see rain outside, and I think it'll be dry in maybe 10 seconds because that’s what the app says.''} P01 developed trust in sleep data after observing frequent alignment with their feelings, and subsequently doubted themselves when the data occasionally misaligned: \textit{``...sometimes even if I think I slept right and maybe that I didn't wake up so much, the Fitbit actually says that my wake up time is more than I expected...I think in that case it is accurate. Maybe I'm not aware.''} 
\textit{Trust in the self} - Many participants (n=14) ignored mismatches because they trusted their feelings. Some (n=3) were disinterested in data concerning sleep and stress, and in related mismatches, being sceptical about data's ability to capture experiences. P12 stated: \textit{``I just disregard the whole thing. If I feel that I would have slept bad...The watch is not gonna tell me that I slept well. I can feel it.''} P11 echoed similar sentiments: \textit{``I wouldn't do anything with it, I think, because I trust more my feeling, I don't trust the watch.''} (P08) \textit{``I think stress is something very subjective and I'm not sure if you can measure it or not...At least for me, I don't think that's gonna work out and that I will ever trust this.''}

\textbf{Explainability and mental models} - System intelligibility affected participants' perceptions of the data-expectation gap. Confusion concerning the emergence of mismatches or the score calculation typically contributed to negative perceptions or ignoring the data. Participants sometimes struggled to understand why their scores contradicted their feelings. P04 wondered, \textit{``is it comparing me to the medium of the people in my area, because then maybe other people my age are sleeping worse than I do?[]Is it because maybe because I was wearing it for only a week, and then it didn't have enough historical data for me to make accurate conclusions?''} Frustration also arose when the system made unrealistic recommendations without explanation. (P13) ~\textit{``...one night it went off and I was like, `why do I need to walk at this moment?'[]Maybe my HR went up very high, that could be possible because that was crazy.''} P02, for instance, felt confused and frustrated when the watch suggested more walking after logging 20K steps at their job: \textit{``Yeah. Fuck you. I never have to go. I walked enough today. So I don't know why it says that.''} Conversely, participants were often more accepting of mismatches when they could explain them. P14, for example, accepted overestimated SQ scores after attributing them to their need for longer than average sleep; \textit{``...if I can change something, the baseline for calculating the score or customize it, which I can do, I think it would match it perfectly.''}

The UI might have also affected the system's capacity to provide clear explanations. Five participants found it complicated to access relevant information. For instance, viewing sleep data required scrolling and pressing a button, while other information was more highlighted: (P04)\textit{``Now you have to dig deep.''} Many even overlooked features like HRV and stress management. P03 also noted the UI made it difficult to retrieve skipped tutorials.

Participants' \textit{background} sometimes influenced their mental models for explaining mismatches. P03, a software developer, and P07, a data scientist, forgave activity classification errors due to this rationalisation: (P07) \textit{``I think it's not a problem because you cannot really identify like 100,000 different classes.''} Similarly, P03 understood the activity classification formula but not sleep detection, resulting in frustration after sleep-related mismatches: \textit{``I was less trustworthy. And I checked it less afterwards.''} Previous knowledge concerning similar features or technologies also impacted mental models and trust. P02 suspected stress detection is related to HR, and knowing that HR is also related to activity made them doubt stress metrics. P08 expressed uncertainty about stress calculation mechanisms, and attributed Garmin's systematic overestimation of their stress to their relatively high resting HR: \textit{``I know that I have a quite higher HR than the average and then somehow it classifies it as stress.''}

\subsection{Key findings}

The second study validated the parameters affecting mismatch perception found in the first study. It also showed the presence of \textit{tension potential} and that the following factors are influential: \textit{emotional state}, \textit{trust}, and \textit{explainability and mental models}.

%% file: Sections/5-Vocabulary.tex
\section{A Vocabulary for the Data-Expectation Gap}

\subsection{The Proposed Taxonomy}

We combined the findings from our two studies into a vocabulary describing the data-expectation gap, grouping themes into three \textit{tension development mechanisms}: \textit{mismatch detection}, \textit{contextualisation} and \textit{personal evaluation}. Our taxonomy's categories can be seen as a set of building blocks where different combinations lead to mismatches with different implications and potential impact, despite having the same underlying data. Below we briefly describe our taxonomy, while more details for each term can be found when it is first introduced in our studies.

\begin{description}[leftmargin=2em,parsep=1em]
    \item \textbf{Data-expectation gap} - a mismatch between detected and expected values related to user behaviours, actions, sensations, beliefs, or feelings, including classification and numeric value detection or estimation, or irrelevant recommendations. A single such instance is termed a \textbf{mismatch}. The data-expectation gap materialises through HDI, as this is where users become aware of mismatches. 
    \item \textbf{Tension development mechanisms} - The \textit{data-expectation gap} may result in tensions based on three \textit{tension development mechanisms}: \textbf{mismatch detection}, \textbf{contextualisation} and \textbf{personal evaluation}.
    \begin{itemize}
 
        \item \textbf{Mismatch detection} - This mechanism describes how users interact with data and detect mismatches with a specific \textbf{type} and \textbf{source of evidence}. This mechanism can contribute to tension development as comparisons with evidence lead to disbelief. 
        \begin{itemize} 
            \item \textbf{Source of evidence} - Detecting a mismatch presupposes comparing the data with another source of evidence. This term, with its subcategories \textbf{\TDataLogic}, \textbf{\TDataMeasure} and \textbf{\TDataFeeling}, specifies this source. The \textit{\TDataLogic} relies on broader knowledge and an assessment of the data's plausibility. The \textit{\TDataMeasure} involves an external source of evidence for comparison. In the \textit{\TDataFeeling}, the evidence is more subjective, including feelings and senses, which may influence the user's certainty in their judgment.
            \item \textbf{Type of mismatch} - This category describes the type of discrepancy between data and expectations. \textbf{Classification issues} include \textit{false detection}, \textit{detection failure} and \textit{partial detection} - accurately identifying an overall state but with errors in details or derivative parameters. \textbf{Value estimation issues} include \textit{high fluctuation}, \textit{overestimation} and \textit{underestimation}. Other issues include \textit{delayed response} and \textit{incorrect location}. 
        \end{itemize}

        \item \textbf{Contextualisation} - This mechanism describes how \textit{contextualisation} contributes to tensions alongside \textit{mismatch detection}. Contextual parameters can interact with the \textit{mismatch detection} mechanism in different ways, shaping the overall experience after a mismatch is detected, or actively driving mismatch detection with the context itself causing disbelief. These parameters are: \textbf{activity}, \textbf{location}, \textbf{temporality}, \textbf{emotional state}, and \textbf{co-experience}.
        \begin{itemize}
            \item \textbf{Activity} and \textbf{location} - The \textbf{activity type} and \textbf{activity intensity}, along with \textbf{environmental affordances and equipment}, can cause or amplify disbelief, cause concern, disappointment or frustration as a result of the data-expectation gap, and cause the data-expectation gap.
            \item \textbf{Temporality} - Temporal factors affecting users' perceptions of mismatches include: \textbf{temporal unit of analysis}, \textbf{moment of mismatch detection}, and \textbf{inconsistency in repeated experiences}. The \textit{temporal unit of analysis} describes how users detect a mismatch by focusing on different aspects - assessing the data \textit{as a whole}, or using \textit{snapshot} or \textit{continuous} assessment. The \textit{moment of mismatch detection} further describes whether users detect the mismatch \textit{during} or \textit{after the activity}, determining whether the detection of the mismatch has real-time consequences for the user's plans. The \textit{inconsistency in repeated experiences} further describes how in repeated interactions with the data, users may detect a mismatch by comparing the data and noticing either score variations in similar situations or identical data in different experiences.  
            \item \textbf{Emotional state} - The user’s emotional state can influence their perception of a \TDataFeeling, with trust in themselves or the data affecting whether their emotions shift from positive to negative or vice versa.
            \item \textbf{Co-experience} - Similar to \textit{activity}, the presence of another individual can cause or amplify disbelief, or cause the data-expectation gap.
        \end{itemize}
        \item \textbf{Personal evaluation} - This mechanism describes how users process the outcome of the interaction between \textit{mismatch detection} and \textit{contextualisation}, and determine the personal meaning of this mismatch in this particular context. Factors contributing to this mechanism include \textbf{personal factors}, \textbf{history of experiences}, and \textbf{explainability}.
        \begin{itemize}
            \item \textbf{Personal factors} - These include the user's \textbf{goals and motivations}, their \textbf{trust} levels, and their \textbf{background}. \textit{Goals and motivations} describe the user's interest in the specific data related to the mismatch. High interest in the data increases the likelihood of tension, as mismatches are more likely to cause frustration. \textit{Trust} in the self and the data describe the user's general tendencies to believe in their feelings or technology. \textit{Background} describes the user's knowledge concerning the data, depending on their profession and any other factors affecting their mental model of the system. 
            \item \textbf{Explainability} - Low system intelligibility negatively affects a user's perception of a mismatch by contributing to confusion about the reason for the mismatch.
            \item \textbf{History of experiences} - This category explains how perceptions of mismatches are influenced by encounters with the data-expectation gap over time. The \textbf{frequency of encounters} with the data-expectation gap describes how the perception of a mismatch is affected by the number of incidents. While one mismatch might sometimes be acceptable depending on its context, the \textbf{co-occurrence of mismatches} can increase the impression that the device is not functioning well. \textit{Co-occurrence of mismatches} includes \textit{between-parameter co-occurrence} - detecting multiple mismatches in different data types - and \textit{within-parameter co-occurrence}, where multiple mismatches with different directions are identified in the same data type (e.g. overestimation and overestimation of the same parameter). The \textbf{cascading effect} describes how detecting a mismatch can cause quality concerns for other data types, or cause other mismatches. In the \textit{parameter-specific cascading effect}, one mismatch leads to further mismatches or mistrust towards related parameters, whereas in the \textit{generalised cascading effect} the mistrust affects all data types. Finally, \textbf{prior encounters} with the data-expectation gap form baseline expectations, making users anticipate certain mismatches or elevating their disappointment. 
        \end{itemize}
    \end{itemize}
        \item \textbf{Tension potential} - This term describes differences in users' probabilities of encountering the data-expectation gap, depending on interaction patterns and variances in the presence of \textbf{potential mismatches} - data points with a probability of being perceived as a mismatch solely based on the data. 
\end{description}

\subsection{Using the Vocabulary}

\subsubsection{Designing Interactions}

Our vocabulary can guide the design of interactions with smartwatch data, helping designers envision different scenarios of experiencing the data-expectation gap and designing appropriate interactions for mitigating emerging tensions. To assist in this process, we provide the following guidelines.

\textit{Create personas and scenarios matching different combinations of building blocks of the vocabulary as follows: Consider how many functions the device has. For each data type related to each function, repeat the following process.}

\begin{enumerate}
    \item Start from the \textit{mismatch detection} mechanism, considering the possible \textit{types of mismatches} for this data type. 
    \item Create pairs including data type, \textit{type of mismatch}, and one or more properties related to the \textit{contextualisation} mechanism (\textit{location}, \textit{activity}, \textit{temporality}, \textit{co-experience}, \textit{emotional state}). Create scenarios based on these pairs:
    \begin{itemize}
        \item \textit{Location} and \textit{activity}: Consider scenarios where the mismatch happens in different combinations of locations and activities: outdoors or indoors, during exercise, rest, everyday activities, or sleep. For each location consider its environmental affordances, and for activities consider the activity type and intensity. For each scenario, consider if it could amplify implausibility, cause the mismatch, or create worry or disappointment from unmet goals.
        
        \item \textit{Temporality}: For each combination of activity and location identified above, imagine: what would be the potential impact if the user identifies the mismatch during or after the activity? What happens if the user checks the data continuously, compared to only checking them intermittently and finding a mismatch in one of these instances while the rest of the data might be accurate? 
        \item \textit{Co-experience}: Consider how the presence of another can affect the experience. Can they trigger or help identify mismatches, experience them too, help their interpretation, alleviate or increase tensions?
        \item \textit{Emotional state}: Consider how the mismatch type can affect a user having a negative or positive emotion - especially when concerning a \TDataFeeling. Consider if there are circumstances where the mismatch can cause worry.
    \end{itemize}
    \item For each scenario, also consider: How can the interaction between context and the \textit{type of mismatch} lead to a mismatch with a different \textit{source of evidence}? How could interaction mechanisms be tailored to different types of evidence that could affect the perception of the data-expectation gap in this specific context?
    \item Enhance these scenarios by considering the building blocks of the \textit{personal evaluation} mechanism: 
    \begin{itemize}
        \item Create personas with different \textit{goals and motivations}, \textit{background} and levels of \textit{trust} in the system and in the data. For the created scenarios, consider:
        \begin{itemize}
            \item How would a user who trusts themselves respond to this scenario, versus one who trusts technology? How can the system create positive experiences for both?
            \item How could the mismatch in this scenario become a barrier for a user with a specific interest related to this function? 
            \item How does the user's background influence their data literacy concerning this data type, and therefore data interpretation? How could the interaction mechanism adapt to users with differences in data literacy levels? Consider this aspect also in the context of systems with different levels of \textit{explainability}.
        \end{itemize}
        \item For the created personas and scenarios, consider how differences in the \textit{history of experiences} might change the experience in each scenario:
        \begin{itemize}
            \item How frequently does the mismatch happen? If this user has already experienced a lot of mismatches, when might a new mismatch be a problem for them? Do these previous experiences concern similar or different data? What if this is the first mismatch they experience but it is so important for them that it might ruin their trust in this or other data types? 
            \item What happens if they have high or low expectations from \textit{prior encounters} with mismatches, in smartwatches or other contexts? 
        \end{itemize}
    \end{itemize}
\end{enumerate}

These questions can highlight often overlooked elements and help designers prototype HDI mechanisms adaptive to these scenarios. For example, consider a smartwatch generating step data, HR, SQ levels, and other metrics. For steps, we can identify \textit{overestimation}, \textit{underestimation}, and \textit{high fluctuation} as possible \textit{types of mismatches}. Then, we can speculate on contextual characteristics for the ``steps-overestimation'' pair. Considering \textit{location} and \textit{activity}, this mismatch could be encountered indoors or outdoors, during exercise, rest, everyday activities, or sleep - all likely instances of a \textit{\TDataLogic}. In contexts not generating steps, the mismatch may increase disbelief, while during exercise, it could cause frustration over inaccurate activity intensity estimates, depending on user goals. Considering \textit{temporality}, detecting a mismatch during an activity where the user aims for a step goal would affect their immediate plans, while detecting the mismatch after it would increase their mistrust. Detecting the mismatch while continuously comparing steps with device output would also let them determine whether this issue persists, compared to doing a momentary check. Considering \textit{co-experience}, users might compare step data when engaged in the same activity with others, leading to a \textit{\TDataMeasure}. The user's \textit{emotional state} does not contribute to this particular mismatch, but it might affect the resulting frustration. 

Now, these potential scenarios will play out differently depending on the \textit{personal evaluation} mechanism. Consider personas with different \textit{goals and motivations}: Marc, interested in tracking his activity during post-surgery recovery, and Susan, who bought the watch for sleep tracking but is occasionally checking other metrics. Evaluating the different pairings between ``steps-overestimation'' and contextual factors, the overestimation of steps would have a greater impact on Marc, disrupting his ability to achieve his goal - especially with high \textit{frequency of encounters}. In contrast, step overestimation would only directly affect Susan's goal if, based on her \textit{background} and data literacy levels, she incorrectly associates step tracking with sleep metrics, reducing her trust in the primary data of interest. Susan's \textit{trust} in this technology, influenced by \textit{prior encounters}, might also shape her assessment. This step-by-step analysis highlights the need for tailored interactions to accommodate as many of these scenarios as possible, or those with a greater impact. For Susan, the design could prioritise building trust by clarifying that step tracking does not affect SQ, while for Marc, the focus could be on personalising his activity plan by finding enjoyable and beneficial activities that minimise step estimation problems. After iterating through these scenarios, we can repeat the same process for the other data types.

\subsubsection{Analysing Interactions}

\begin{figure*}[h]
    \centering
    \includegraphics[width=1\textwidth]{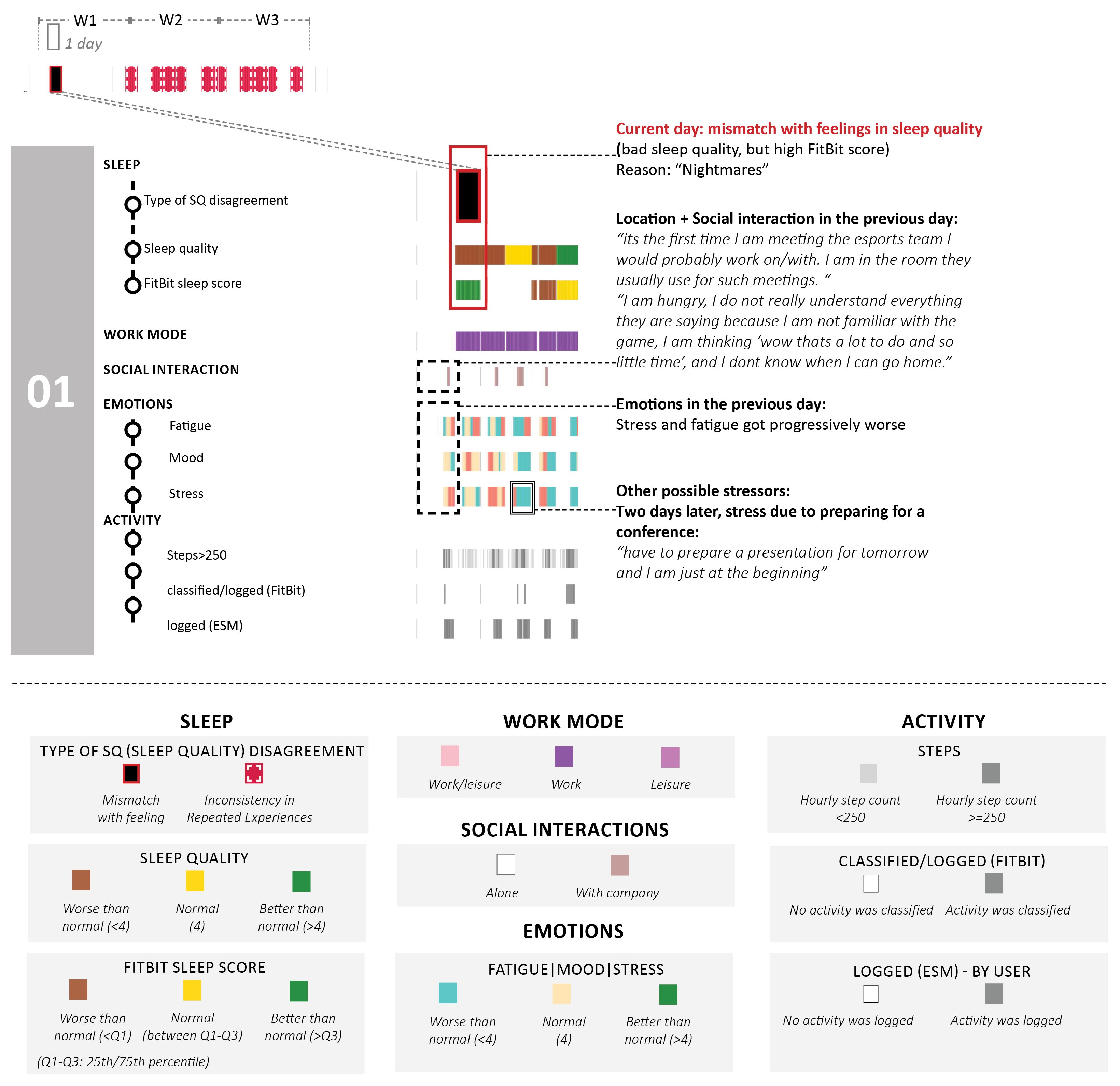}
    \caption{P01's negative experience of a mismatch in SQ data. The figure shows a snapshot of P01's Fitbit and ESM data, focusing on 5 days displayed as a timeline. We show all data that may have affected P01's experience with this mismatch, annotated with their notes from the ESM app.}
    \label{fig:P01SleepDetails}
\end{figure*}

Our vocabulary can also aid in analysing experiences with the data-expectation gap in smartwatches, using its building blocks to identify commonalities and differences between experiences. For example, users' experiences might share similar \textit{contextualisation} mechanisms, but differ in \textit{personal evaluation} mechanisms, with some users accepting mismatches and others not due to differing \textit{goals and motivations} or \textit{history of experiences}. Mapping these commonalities and differences in case studies will help clarify the relationships between mechanisms and building blocks, advancing theory development on the data-expectation gap and the emergence of tensions in interactions with mismatches.

To demonstrate this use, we used our vocabulary to analyse two incidents from our field study. The first incident concerns a case where despite sleeping more than usual, P01 rated their SQ much lower than their median due to nightmares, while Fitbit rated it highly (Figure ~\ref{fig:P01SleepDetails}). This was, therefore, a \textit{\TDataFeeling} related to SQ \textit{overestimation}, concerning the \textit{mismatch detection} mechanism. Analysing the \textit{contextualisation} mechanism, we began by examining the \textit{emotional state}. Despite feeling calm that morning, the previous evening was stressful, in an unfamiliar setting, with late-night work and anxiety about a new project and conference preparations. These feelings contributed to a negative emotional state which might have influenced the experience. The \textit{activity} did not play a role here, and the unfamiliar \textit{location} only indirectly influenced their negative emotions. Moving to the \textit{personal evaluation} mechanism, while this was a strongly negative experience as evident in P01's recollections, they still found the sleep statistics overall relatively accurate and helpful in explaining their tiredness. This was because of their \textit{goals and motivations}, as P01 has a condition affecting their sleep, making them particularly interested in tracking it. P01's condition also affects their \textit{trust} in themself and their perception of their bodily state, as it can amplify painful sensations. Figure ~\ref{fig:disagreements_sleep} shows that besides this mismatch, P01 also had instances of strong alignment between reported and perceived SQ, in days of good and bad sleep. These positive experiences and lack of prior exposure to better devices - as this was their first device, regarding the theme of \textit{prior encounters} - might have contributed to their assessment.

\begin{figure*}[h]
    \centering
    \includegraphics[width=1\textwidth]{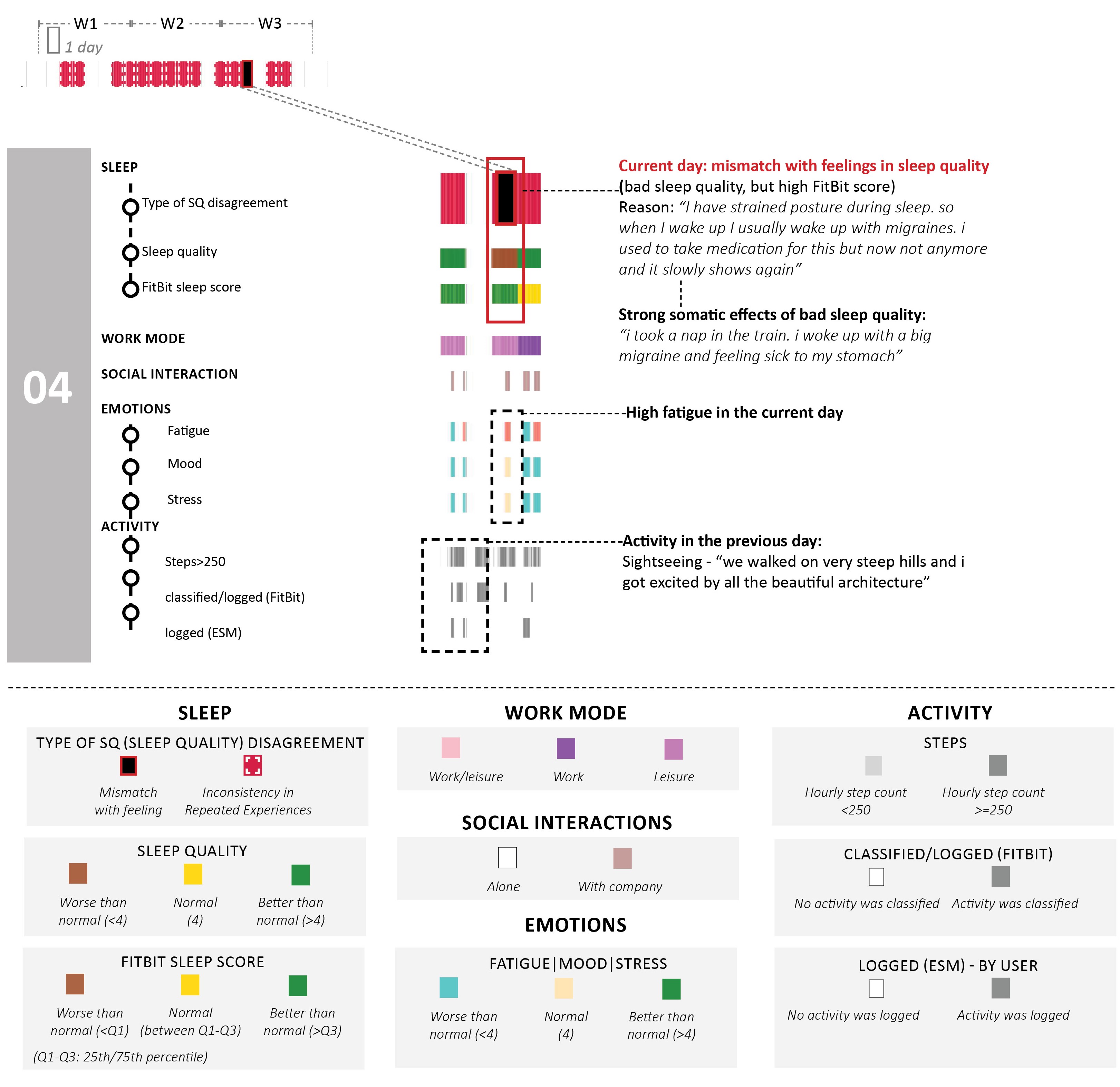}
    \caption{P04's negative experience of a mismatch in SQ data, visualised following the same approach as Figure~\ref{fig:P01SleepDetails}.}
    \label{fig:P04SleepDetails}
\end{figure*}

Similarly, Fitbit gave a sleep score of 90 to P04 (Figure ~\ref{fig:P04SleepDetails}) who gave a worse than usual rating after a relatively long sleep. The \textit{mismatch detection} mechanism here is similar to the previous example, as this was another example of \textit{\TDataFeeling} related to SQ \textit{overestimation}. Analysing the \textit{contextualisation} mechanism, we found that on the same day, P04 mentioned sleep-related issues related to strained posture in the ESM log. This incident was during a trip abroad, and they were very fatigued after having walked much more than usual on the previous two days, based on their step data; they woke up feeling sick, and later napped on the train. The Fitbit score was not simply different from their feeling; strong somatic effects affected the sensual layer of the experience~\cite{mccarthy2007technology}, adding to their \textit{emotional state}. The \textit{location}, \textit{activity} and \textit{co-experience} here set the experience in a foreign environment with considerable activity. Analysing the \textit{personal evaluation} mechanism, we found that sleep tracking was related to P04's \textit{goals and motivations} as they had history with sleep-related issues due to strained posture, and wanted to improve their SQ. Their recollection of the experience highlights their \textit{trust in the self} and the resulting frustration; \textit{``I always put my feelings about the data and if I feel bad, but the sensor says like feel good, it feels a bit betraying...It feels like I'm going and saying, I have psoriasis. And the doctor says, I don't see anything.''} Overall, P04 mistrusted the sleep data, possibly due to \textit{prior encounters}: they had fewer experiences of strong alignment between perceived and detected SQ compared to P01 (Figure ~\ref{fig:disagreements_sleep}) and unmet expectations due to previously having a more accurate watch.

This analysis shows how our vocabulary can be used as an analytical tool to identify differences and similarities in experiences with the data-expectation gap. For example, P04's and P01's experiences both concern a \textit{\TDataFeeling} related to SQ \textit{overestimation} but involve unique combinations of contextual circumstances and personal factors that shape each experience differently. We propose using the vocabulary as an analytical tool in such a way that allows this individuality - which is a key theme in the TaE framework~\cite{mccarthy2007technology} - to surface, while also highlighting commonalities.

%% file: Sections/6-Discussion.tex
\section{Discussion}

Overall, our findings highlight the complexity of experiencing the data-expectation gap in interactions with smartwatches. The vocabulary we generated, which is our main contribution, provides a new terminology that describes nuances in experiencing the data-expectation gap~\cite{Consolvo2008Ubifit}. Here we further discuss our vocabulary and findings, along with implications for HDI.

\subsection{Connection With Previous Studies}

Our vocabulary consolidates previously fragmented information on the emergence and perceptions of mismatches into a single taxonomy, enriched by our new findings - particularly concerning the importance of context. Understanding the contextual factors surrounding experiences of mismatches is key to understanding their meaning, as the same mismatch type can have different implications when paired with different contextual factors. Our findings provide a structured way to analyse how situated objectivity~\cite{Pantzar2017SituatedObjectivity} manifests in interactions with data~\cite{Pantzar2017SituatedObjectivity}.

Our findings also enrich existing knowledge on the data-expectation gap (Table \ref{tab:Review_table}). Similar to Shih et al., and other studies~\cite{mackinlay2013fitbitaccuracy, Consolvo2008Ubifit,shih2015use}, we found that \textit{prior encounters} shape expectations about device capabilities. Among sources of evidence users evaluate to identify mismatches, the \textit{\TDataMeasure} and \textit{\TDataLogic} connect to the work of Yang et al.~\cite{Yang2015Accuracy}, while the \textit{\TDataFeeling} connects to Ding et al.'s work on the subjectivity of stress tracking~\cite{Ding2021Stress}. Concerning \textit{co-experience}, while other studies have suggested that co-experiencing data could help alleviate tensions from mismatches~\cite{Ding2021Stress,Jiang2023Intimasea}, we showed that the mechanisms related to this phenomenon are complicated, as social interactions can also trigger the detection of a mismatch. Our findings concerning the relationship between data interpretation and trust in the self and technology align with Rapp and Cena~\cite{rapp2016personal}, and Lomborg et al.~\cite{lomborg2018temporal}. We further discuss the topic of trust in the next section, along with the role of emotions in the perception of mismatches.

\subsubsection{The Development of Mistrust in Experiences With the Data-Expectation Gap}

Prior research has linked accuracy to trust in smartwatches~\cite{Michaelis2016Wearable}. Our findings further show that multiple parameters can contribute to the development of mistrust when experiencing a mismatch. Trust in the data is often difficult when disbelief is high. Our analysis of the \textit{history of experiences} indicates that trust issues may arise rapidly after one incident for some, and after repeated mismatches for others. Changes in trust levels over time highlight its dynamic nature~\cite{cho2015trustsurvey}. Disappointment from accuracy expectations, as found in our analysis of \textit{prior encounters}, may lead to general mistrust towards similar technologies, as mistrust from negative experiences of disconfirmed expectations can cascade to other experiences~\cite{darke2010disconfirmtrust}. Users' mental models, affected by system intelligibility~\cite{Abdul2018XAI}, user background, and differences in data literacy~\cite{wolff2016dataliteracy}, are also utilised in trust-related assessments. Users sometimes lack clarity on how algorithms work in personal tracking~\cite{Yang2015Accuracy} and might discard correct outputs or believe faulty data~\cite{bansal2019beyondaccuracy}. Trust is also shaped by attitudes towards technology and belief in oneself. This was shown in the field study, where some users trusted sensor data over their feelings, while others were sceptical of specific features or technology.

Overall, our work shows that mistrust is affected by experiential qualities of mismatches, and personality, background knowledge, and beliefs. Users distrustful after one mismatch have less opportunity to regain trust than those frustrated by multiple incidents. This difference should be considered when designing HDI mechanisms, along with increasing data literacy and system intelligibility, as these elements affect mental models and trust. Minor mismatches can also affect the user's mental model and impact future expectations and attitudes towards technology; therefore, it is important to consider the broader role that experiencing a mismatch might have on shaping trust and attitudes towards technology, beyond momentary reactions.

\subsubsection{Influence of Algorithmic Output on Emotions}

The field study showed that users' emotional states affect their perception of data contradicting their feelings. In \cite{Ding2021Stress}, users were confused when encountering mismatches at both ends of stress levels, leading to mistrust, and for some, device abandonment. In our study, while users trusted themselves when certain about their feelings, data suggesting a worse state sometimes caused worry. This finding aligns with research showing that encountering data with negative connotations can lead to negative experiences~\cite{Epstein2016AbandonTracking, Howell2016Hint}. Some users also felt better after encountering more positive data, but a few felt their negative emotions were invalidated. This contradicts previous research showing that positive feedback related to emotions can improve the perception of an experience~\cite{Hollis2018DataEmotions}. This discrepancy could be due to participants’ certainty in their assessment. Overall, our findings highlight the complexity of the relationship between algorithmic output, interpretation, and emotions in subjective data. Given the small sample in our second study, we encourage future research to clarify the mechanisms driving emotional responses to emotion-related data.

\subsection{Moving Beyond Traditional Error metrics}

While our study focused on the data-expectation gap in smartwatch use, mismatches between data and expectations occur in various contexts, from office wellbeing perceptions~\cite{Brombacher2024Clicker} to communication with voice assistants~\cite{palanica2019voice}. Data errors are typically assessed using metrics like precision and recall~\cite{hossin2015accuracyevalmetrics} - which align with terms from our \textit{type of mismatch} category such as \textit{false detection}. While such metrics are essential for evaluating sensor or algorithm performance, our findings highlight that they do not capture key experiential qualities of the data-expectation gap, which can change a mismatch's implications. Our vocabulary instead highlights experiential qualities that can help UX designers understand nuances in data and artificial intelligence (AI) failures, and mitigate tensions when prototyping data-enabled systems~\cite{Yang2020DesigningHumanAIInteraction}. 

While this vocabulary was based on smartwatch use, it can serve as a foundation for examining the data-expectation gap in other contexts. We anticipate that the need for such research will grow in the coming years, as we increasingly collaborate with AI, and it is critical to understand how to handle AI errors in these collaborations. Recent studies have explored how people respond to different types of AI errors; e.g. concerning the error magnitude and rate of occurrence~\cite{Erlei2024ErrorsAI} and error timing~\cite{Kim2023AlgorithmsErr}. Our vocabulary can advance this research proposing an experience-informed approach to understanding data errors. For example, consider interacting with a Large Language Model (LLM). The data-expectation gap would emerge upon a semantic discrepancy between the expected and produced LLM output. Contextual factors like activity, location, co-experience, and temporality, can change the interpretation and impact of such a mismatch in similar ways as in our case study. Some terms, of course, will disappear in certain contexts, or gain new interpretations, while new terms might emerge. For example, the \textit{\TDataMeasure} could be interpreted as a case where the user compares the text with another text corpus, while the \textit{types of mismatches} would need to incorporate new terms like factual errors or incomplete information. An experience-based approach to analysing mismatches, like ours, can be valuable for tailoring HDI mechanisms to different error scenarios. We hope that our work will pave the way towards complementing traditional analyses of errors with approaches such as our vocabulary.

\subsection{Towards Tension Mitigation Mechanisms in HDI}

While our work showed the breadth and depth of issues related to the data-expectation gap, going forward with this knowledge is still a challenge, due to the lack of research towards designing for tensions in HDI. Algorithmic accuracy alone cannot resolve this; even at high levels, it will never reach 100\%. The data-expectation gap will, therefore, always exist to some degree. Since we cannot eliminate it, the best we can do is to understand it and develop mechanisms that mitigate the resulting tensions. These mechanisms can draw from existing approaches like making algorithmic systems explainable, intelligible and transparent~\cite{Yang2015Accuracy} and outlining the device's technical capabilities~\cite{shih2015use}. Another suggestion is to enable user feedback, which can drive customisation~\cite{Yang2015Accuracy} and algorithm personalisation~\cite{Harrison2015Barriers,Yang2015Accuracy}. Negative feedback can also take the form of contestation, particularly when HDI involves emotional biosensing and inferences based on subjective data~\cite{Howell2018biosensing}. A different approach could also be to increase self-reflection, for example through microboundaries~\cite{Cox2016Frictions} - small obstacles providing opportunities for reflection, to induce more mindful interaction through friction.

While these approaches offer opportunities for addressing the data-expectation gap, many questions remain: When do users prefer explanations, giving feedback, or self-reflection, after experiencing a mismatch? When is negotiation with the algorithm or collaborative human-AI sensemaking of mismatches appropriate? Could we envision tangible tension mitigation mechanisms where the building blocks of our vocabulary are translated into properties like form and materiality?  We hope our work inspires future research in these areas.

\subsection{Limitations and Future Work}

Our study also has limitations. We did not include any users with disabilities in our sample, and Quantified-Selfers~\cite{Choe2014QSPractices}, who may have different behaviours and attitudes. The data we analysed also have limitations; in interviews, recollections may be prone to cognitive bias, while the ESM data also had gaps. These limitations were addressed by combining these sources. While we cannot guarantee the truthfulness of the online review data, the findings aligned with the field study findings, supporting their validity. In the field study, users were given the equipment; had they purchased it, their experiences might have been coloured by higher satisfaction or regret. We also did not explicitly ask the participants to log all the experienced mismatches during the study. This affected the identification of experienced mismatches in the field data but reduced the possibility of users being overly attentive to mismatches. Furthermore, while our online review study covered diverse smartwatches, the field study focused on a specific device. Future work can validate our findings by interviewing owners of different models. We also did not investigate how aesthetics and interface design might affect the experience of mismatches. Future work should look further into these aspects. Finally, future work could investigate how the data-expectation gap is experienced in ecologies of devices beyond a single smartwatch.

%% file: Sections/7-Conclusion.tex
\section{Conclusion}

In this paper, we analysed encounters with the data-expectation gap from online product reviews and a field study. We distilled our findings into a vocabulary describing the experiential qualities of these encounters and their link to the emergence of tensions between the user and technology. These qualities include factors related to \textit{contextualisation} of the data-expectation gap - temporality, activity, location, co-experience, and emotions - and \textit{personal evaluation} - history of encounters, users' background, goals and motivations, trust in oneself and the data, system explainability, and other factors affecting the escalation of importance of a mismatch. We hope this vocabulary serves as the foundation for future research on designing tension-aware HDI mechanisms.

%% file: Sections/Appendix.tex
\section{Appendix: Online review study}

\subsection{Selection of online reviews} \label{Appendix: Review selection}

We sourced product reviews from Amazon, covering nine regions: USA, UK, Germany, India, Japan, Mexico, Brazil, Saudi Arabia, and UAE. We chose five smartwatches - Fitbit Sense and Sense 2, Garmin Fenix 6 and Forerunner 235, Xiaomi Redmi Watch 2 Lite - based on their release within the last five years, their wide range of supported features (having sleep, HR, and activity tracking as a baseline), and their diversity in pricing (between 60-450\$ in Amazon US during our search). We occasionally bypassed reviews when a site had fewer than ten reviews for a model. Reviews were automatically translated by Amazon into English where needed. 

We initially screened 5345 reviews, first reading the reviews manually, and then scanning again the text for the following terms and variations of the root words to check for any ommissions: \textit{accurate, inconsistent, reliable, fault, correct, precise, error, mistake, measure, count}. 14\% (745) of these screened reviews mentioned accuracy issues. From this 14\%, we chose 200 reviews for further analysis. The 200 reviews were selected by including from each brand the first 50 reviews that included information related to the context (e.g., the activity or location related to detecting a mismatch). This inclusion criterion was the reason that we excluded the Xiaomi Redmi Watch 2 Lite from the final sample, as we only found 12 reviews with contextual information. 

\subsection{Protocol for coding the data from online reviews} \label{Appendix: review coding protocol}
We looked for themes indicating different types of mismatches in terms of the data types they concerned, the process of identifying the mismatch, the relationship to known performance metrics (e.g. classification accuracy), the relationship with themes from the TaE~\cite{mccarthy2007technology} framework (e.g. the compositional, sensual, emotional, and spatio-temporal threads that characterise experience), and participants' attitudes towards the data-expectation gap. We also used a context categorisation for mobile UX analysis~\cite{Korhonen2010MobileUXwithContext} as inspiration for analysing the contextual parameters influencing perceptions of the data-expectation gap. While we purposefully searched the data for references relevant to these topics, we also generated new codes and themes during this process. Therefore, our analysis included a mix of inductive and deductive coding. 

\section{Appendix: Field study}

\subsection{Fitbit Inspire 3 characteristics} \label{Appendix: Fitbit characteristics}

Fitbit Inspire 3 supports the measurement of steps, HR, activity intensity, distance, calories, floors, SQ and sleep duration, resting HR, activity classification (for the following activities: walk, run, outdoor bike, elliptical, sports, aerobic workout, swimming~\cite{Fitbit2023}), and has a stress management score. Some features are specifically measured during the night (HRV, breathing rate, skin temperature). Some additional features, such as the daily readiness and a detailed sleep profile, are only accessed with a premium subscription, which we did not provide to our participants. 

\subsection{ESM protocol} \label{Appendix: ESM protocol}

\begin{figure*}[t]
\centering
\begin{minipage}{1\textwidth}
  \centering
  \captionsetup{width=1\linewidth}
  \includegraphics[width=\textwidth]{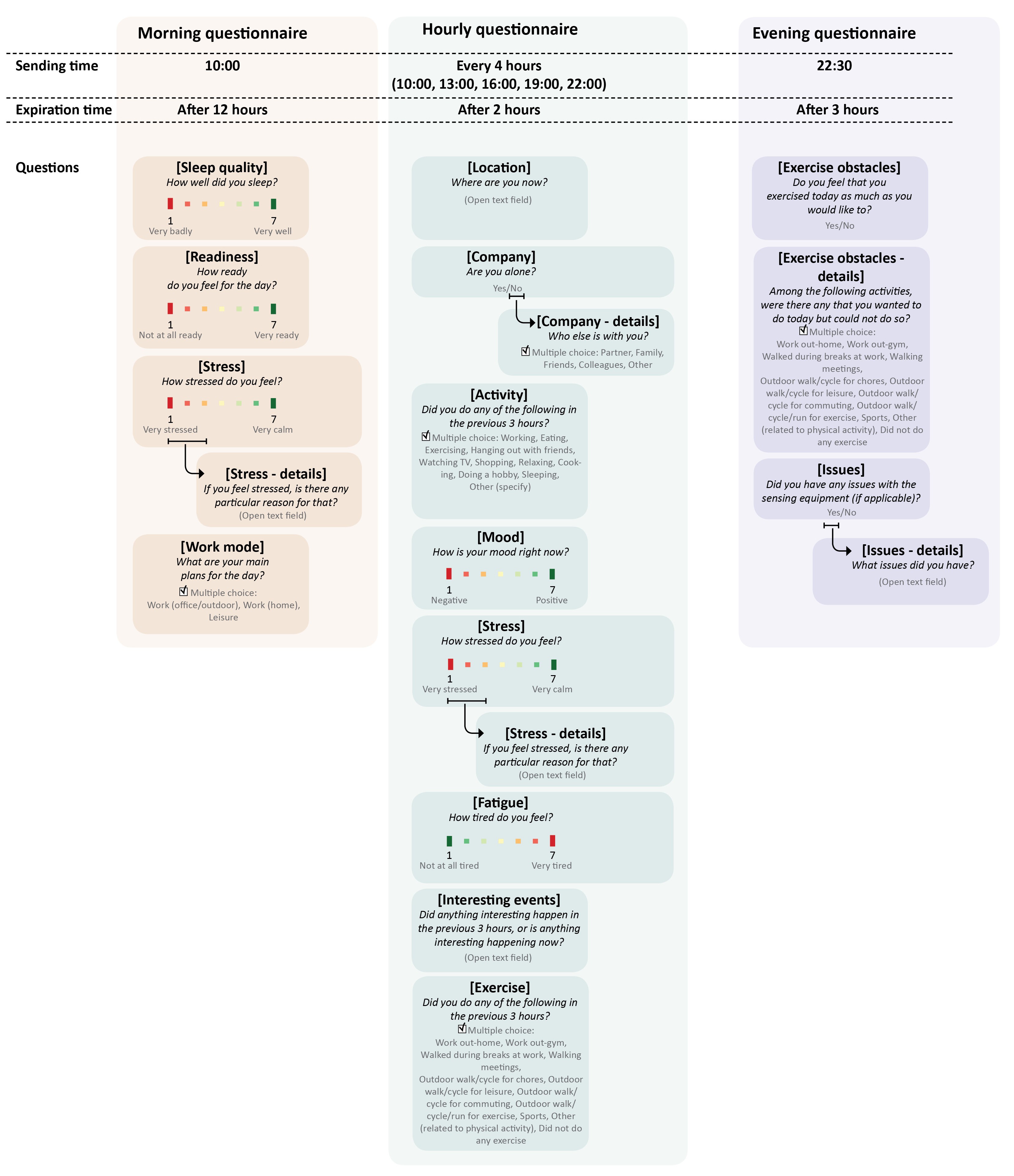}
  \captionof{figure}{The ESM protocol.}
  \label{fig:ESM protocol}
\end{minipage}%
\vspace{-1.5em}
\end{figure*}

We chose to use a 7-point Likert item for our questions related to subjective stress and SQ instead of the 100-item scale used by Fitbit, considering that scales with more than 10 response options may be less stable considering test-retest reliability~\cite{Preston2001scales}, and that it would be best to use the same scale for all questions involving rating, to avoid any confusions. The labels that we used for logging activities in our ESM protocol were similar enough to the activity labels in Fitbit to allow comparisons between the two, but included more contextual descriptors (e.g. ``outdoor walk for leisure'') to enhance our understanding of the circumstances. 

\subsection{Interview protocol} \label{Appendix: Interview protocol}

Note: Questions starting with (++) are those that were primarily analysed in the interview data analysis of Study 2. Information from questions starting with (+) was also considered when relevant.

\begin{itemize}
    \item \textbf{General questions and motives}
    \begin{itemize}
        \item (++)Age and occupation
        \item (++)Previous experience with similar equipment
        \item Reason for participating
    \end{itemize}
    \item \textbf{Personality and experiences during the experiment}
    \begin{itemize}
        \item Would you describe yourself as introspective or not?
        \item Would you describe yourself as optimistic or pessimistic?
        \item Would you describe yourself as a generally happy person?
        \item Would you describe yourself as a generally stressful person?
        \item (+)Do you generally sleep enough?
        \item (+)Is your sleep schedule regular or not?
        \item (+)Is your sleep quality good (satisfying) or not?
        \item (+)What were your experiences with sleep and stress during the monitoring period?
    \end{itemize}
    \item \textbf{Sensor data collection}
    \begin{itemize}
        \item (+)What did you think of your experiences with using the Fitbit? Was there anything that you found interesting, useful or annoying?
        \item (+)Did you check the different scores, such as sleep quality, steps, stress, activity? Were you particularly interested in any of these?
        \item (+)Did you check the application?
        \item (Introduce the scenarios related to tensions)
        \begin{itemize}
            \item (++)Did you have any experiences related to this scenario? If not, what would you think about this?
            \item (++)How would you like to deal with it? (introducing the following concepts: manual editing, feedback, getting an explanation)
        \end{itemize}
        \item (+)Did the use of the Fitbit make you notice anything about your behaviour?
        \item (+)Did you miss anything from the application?
        \item (+)Would you like to see anything more in the application to be able to understand the data?
        \item (+)When were you taking off the device?
        \item (+)Did you notice any recommendations related to sleep or activity? If yes, what did you do?
    \end{itemize}
    \item \textbf{ESM data collection}
    \begin{itemize}
        \item What did you think of the m-Path app? Was there anything that you found interesting, useful or annoying?
        \item Did you skip any questions? If yes, why?
        \item If in a question you had to give the same information as before, how did you feel when you had to respond again while there was no change in your status?
        \item Would you like to add any other data to the questionnaire? If you could add your own questions about anything, what would you add?
        \item How did you feel about the “Interesting” question (if anything interesting happened in the previous hours)? If you did not type anything there, why was that?
        \item What did you think about the response frequency? How would an ideal prompt schedule look like for you?
        \item Did the use of the ESM app make you notice anything about your behaviour?
        \item Did you check the graphs of your data in the application?
        \item Would you like to see anything more in the application to be able to understand the data?
        \item Would you like to combine the subjective data with any data from Fitbit?
        \item In some questions, you were able to see your last response. Did you notice it? If yes, how did you feel about this?
        \item Anything else that you would like to see in the application?
        \item (+)How would you interpret a ‘4’, ‘1’ or ‘7’ in the scale of stress and the other parameters? What would be a situation where you would give these scores?
        \item How did you interpret the question about "being alone"?
    \end{itemize}
    \item \textbf{Reflection on both modes}
    \begin{itemize}
        \item Did you find any of the two modes more valuable than the other?
        \item Would you consider using any of these technologies and apps in the future?
        \item What would you like to see from the data we collected? How much data would an ideal dashboard have, and how frequently would you visit it?
        \item If we used mechanisms such as machine learning to infer some of the contextual insights, instead of asking you via the ESM app, what would you think about this?
    \end{itemize}
\end{itemize}

\subsection{Aligning Fitbit activity data with ESM logs} \label{Appendix: Fitbit-ESM alignment}
To align Fitbit activity data with ESM logs, we linked the activities that Fitbit can classify to activities that participants logged in the ESM data, wherever applicable (e.g. ``walk for commuting'' was matched to ``walk''). Automatic detection requires doing the activity at least for 15 minutes, a filter that participants can change~\cite{Fitbit2023}. We could not check if participants changed this filter or not in the data. 

Mismatches in paired numeric data were discerned by identifying a) data points that were in opposing quartiles based on the IQR (e.g. subjective SQ<Q1 and Fitbit sleep score>Q3, where Q1=25th percentile, Q3=75th percentile), and b) repeated experiences where the score was stable in one dimension while varying considerably - more than the interquartile range - in the other dimension (e.g. Fitbit sleep score was 75 in more than one day, while the subjective SQ was 1-7). We experimentally determined this protocol, aiming to identify a) instances of strong agreement or disagreement, and b) instances of \textit{inconsistency in repeated experiences}, which a few participants recalled in the interviews.